\documentclass[a4paper,fleqn,usenatbib]{mnras}

\usepackage[T1]{fontenc}
\usepackage{ae,aecompl}

\usepackage{graphicx}   
\usepackage{amsmath}    
\usepackage{amssymb}    
\usepackage{txfonts}

\def\aj{{AJ}}%
\def\apj{{ApJ}}%
\def\apjl{{ApJ}}%
\def\apjs{{ApJS}}%
\def\aap{{A\&A}}%
\def\pasa{{PASA}}%
\def\nat{{Nature}}%
\def\mnras{{MNRAS}}%

\title[Rate of Superluminous Supernovae]{The Volumetric Rate of Superluminous Supernovae at $z\sim1$}
\author[S. Prajs et al.]{S. Prajs$^{1}$\thanks{E-mail:
S.Prajs@soton.ac.uk}, M. Sullivan$^{1}$, M. Smith$^{1}$, A. Levan$^{2}$, N. V. Karpenka$^{1}$, T. D. P. Edwards$^{3}$,
\newauthor{C. R. Walker$^{4}$, W. M. Wolf$^{5}$, C. Balland$^{6}$, R. Carlberg$^{7}$, A. Howell$^{8}$, C. Lidman$^{9}$,}
\newauthor{R. Pain$^{6}$, C. Pritchet$^{10}$, V. Ruhlmann-Kleider$^{11}$}\\
$^{1}$Department of Physics and Astronomy, University of Southampton, Southampton, SO17 1BJ, UK\\
$^{2}$Department of Physics, University of Warwick, Coventry, CV4 7AL, UK\\
$^{3}$GRAPPA Institute, University of Amsterdam, Science Park 904, 1090 GL Amsterdam, Netherlands\\
$^{4}$Jodrell Bank Centre for Astrophysics, School of Physics and Astronomy, The University of Manchester, Manchester M13 9PL, UK\\
$^{5}$Kavli Institute for Theoretical Physics, University of California, Santa Barbara, CA 93106, USA\\
$^{6}$LPNHE, CNRS-IN2P3 and University of Paris VI \& VII, F-75005 Paris, France\\
$^{7}$Department of Astronomy and Astrophysics, University of Toronto, 50 St. George Street, Toronto, ON M5S 3H8, Canada\\
$^{8}$Las Cumbres Observatory Global Telescope, 6740 Cortona Dr., Suite 102, Goleta, CA 93111, USA\\
$^{9}$Australian Astronomical Observatory, North Ryde, NSW 2113, Australia\\
$^{10}$Department of Physics and Astronomy, University of Victoria, P.O. Box 3055, Victoria, BC V8W 3P6, Canada\\
$^{11}$CEA, Centre de Saclay, irfu SPP, 91191 Gif-sur-Yvette, France}
\begin{document}
\label{firstpage}
\pagerange{\pageref{firstpage}--\pageref{lastpage}}
\maketitle

\date{\today}

\begin{abstract}
  We present a measurement of the volumetric rate of superluminous
  supernovae (SLSNe) at z$\sim$1.0, measured using archival data from
  the first four years of the Canada-France-Hawaii Telescope Supernova
  Legacy Survey (SNLS).  We develop a method for the photometric
  classification of SLSNe to construct our sample. Our sample includes
  two previously spectroscopically-identified objects, and a further
  new candidate selected using our classification technique. We use
  the point-source recovery efficiencies from
  \citet{2010AJ....140..518P} and a Monte Carlo approach to calculate
  the rate based on our SLSN sample.  We find that the three
  identified SLSNe from SNLS give a rate of
  $91^{+76}_{-36}$\,SNe\,Yr$^{-1}$\,Gpc$^{-3}$ at a volume-weighted
  redshift of $z=1.13$. This is equivalent to
  2.2$^{+1.8}_{-0.9}\times10^{-4}$ of the volumetric core collapse
  supernova rate at the same redshift. When combined with other rate
  measurements from the literature, we show that the rate of SLSNe
  increases with redshift in a manner consistent with that of the
  cosmic star formation history. We also estimate the rate of
  ultra-long gamma ray bursts (ULGRBs) based on the events discovered
  by the \textit{Swift} satellite, and show that it is comparable to
  the rate of SLSNe, providing further evidence of a possible
  connection between these two classes of events. We also examine the
  host galaxies of the SLSNe discovered in SNLS, and find them to be
  consistent with the stellar-mass distribution of other published
  samples of SLSNe.
\end{abstract}


\section{Introduction}
\label{sec:introduction}

Superluminous supernovae (SLSNe), defined as events with an absolute
magnitude brighter than $-$21 ($M<-21$), are a new puzzle in the study
of supernovae \citep{2012Sci...337..927G}. They appear 50-100 times
brighter than normal supernova events, and form at least two distinct
classes: SLSNe-II, which show signatures of interaction with
circumstellar material (CSM) via hydrogen and other lines
\citep{2007ApJ...659L..13O,2007ApJ...666.1116S,2011ApJ...735..106D},
and SLSNe-I (or SLSNe-Ic), which are hydrogen poor
\citep{2011Natur.474..487Q}.  While SLSNe-II may naturally be
explained as an extension of the fainter type IIn supernova events,
the power source behind SLSNe-I remains a subject of debate.

The most popular model in the literature to explain SLSNe-I involves
energy input from the spin-down of a newly-formed magnetar following a
core-collapse supernova
\citep{2010ApJ...717..245K,2010ApJ...719L.204W,2013ApJ...770..128I},
although alternative models involving pulsational pair instability
supernovae \citep{2007Natur.450..390W,2015ApJ...814..108Y} or
interaction with a hydrogen-free CSM
\citep{2011ApJ...729L...6C,2013ApJ...773...76C,2015arXiv151000834S}
have also been proposed. Additional clues are also provided by the
environments in which SLSNe-I occur: predominantly vigorously
star-forming and low-metallicity dwarf galaxies
\citep{2014ApJ...787..138L,2015MNRAS.449..917L}. This preference for
low-metallicity environments is supported by the modelling of the
SLSN-I spectra, which favours a fairly low metal abundance
\citep{2016MNRAS.458.3455M}.

Of particular note is the rarity of SLSN-I events. It took many years
for the first events to be identified as such
\citep{2007ApJ...668L..99Q,2009ApJ...690.1358B}, and for the class to
be recognised \citep{2011Natur.474..487Q}, in part due to their blue
and relatively featureless optical spectra. Even after several years
of study, only around 25 well-observed SLSNe-I exist \citep[e.g., see
compilations
in][]{2014ApJ...796...87I,2015MNRAS.449.1215P,2015MNRAS.452.3869N}.
Initial estimates placed the rate of SLSNe-Ic at less than one for
every 1000 core collapse supernovae \citep{2011Natur.474..487Q}, and more
recent studies are broadly consistent with this
\citep{2013MNRAS.431..912Q,2015MNRAS.448.1206M}.  However, there has
been no direct measurement of the SLSN-I rate for a well-controlled
optical transient survey. Such a measurement can provide constraints
on progenitor models, as there must, at the very least, be a
sufficient number of any putative progenitor system to produce the
observed SLSN rate. Furthermore, if the strong preference for a young,
low-metallicity environment reflects a real physical effect, any
evolution in the SLSN rate with redshift should also track the cosmic
star-formation and metal enrichment history of the Universe, and the
underlying evolving populations of galaxies.

In this paper, we present such a measurement of the volumetric rate of
SLSNe using data taken from the Supernova Legacy Survey (SNLS), part
of the Canada-France-Hawaii Telescope Legacy Survey (CFHT-LS). SNLS
observed four square degrees of sky for six months per year over the
course of five years, with a primary goal of locating and studying
type Ia supernovae (SNe Ia) to measure the equation-of-state of dark
energy \citep{2006A&amp;A...447...31A,2011ApJ...737..102S}. A
by-product of this search was a deep, homogeneous catalogue of optical
transients down to a limiting magnitude of $i\sim24$
\citep{2010AJ....140..518P}, with more than 500 optical transient
spectra, including two confirmed SLSNe-I \citep{2013ApJ...779...98H}.
This, combined with a significant amount of ancillary redshift
information in the search
fields \citep[e.g.,][]{2006A&amp;A...457..841I,2007ApJS..172...70L,2013A&A...559A..14L, 2013PASA...30....1L},
makes it a perfect controlled dataset for the study of supernova rates
\citep{2006AJ....132.1126N,2009A&A...499..653B,2012AJ....144...59P},
including SLSNe.

This paper is presented as follows. In Section \ref{sec:Defining} we
describe our model for selecting SLSNe, in terms of a magnetar model
that can reproduce the optical luminosity evolution of these events.
Section~\ref{sec:Selecting} introduces the SNLS dataset, and discusses the methods for identifying and
selecting SLSNe from its archive.  Section \ref{sec:MC} focuses on the
Monte Carlo method used to compute the rate of SLSNe using detection
efficiencies measured from the SNLS data. The results are compared
against other SLSN rate measurements in Section \ref{sec:Discussion},
and we summarize in Section~\ref{sec:Summary}.  Throughout, we assume
a Hubble constant of $H_0=70$\,km\,s$^{-1}$\,Mpc$^{-1}$ and a flat
$\Lambda$CDM universe with $\Omega_M = 0.3$, and all magnitudes are
given in the AB photometric system.

\section{Selecting Superluminous Supernovae}
\label{sec:Defining}

Our first task is to develop a method for the photometric selection of
optical transients that will enter our SLSN sample. Although SLSNe are
often defined as supernovae with an absolute magnitude in the
$u$-band, $M_u$, of $M_{u}<-21$ \citep{2012Sci...337..927G}, we do not
use this definition for two reasons. The first is that there are now
several examples in the literature of events that are
spectroscopically similar to SLSNe-I, but that do not pass this formal
threshold.  Examples of these include events such as DES13S2cmm
\citep{2015MNRAS.449.1215P} and LSQ14mo \citep{2015ApJ...815L..10L}.
The second reasons is the recent discovery of new classes of fast and
luminous transients \citep{2016ApJ...819...35A} with luminosities
similar to SLSNe, but with a faster light curve evolution and
different spectroscopic types.  Therefore, in place of an arbitrary
magnitude cut, we use a photometric classification approach built
around a simple analytical model that provides a good fit to a sample
of confirmed SLSNe-I.

There are two main ingredients that we need for our SLSN modelling: an
underlying model for the time-dependent bolometric luminosity of a
SLSN-I, and a spectral energy distribution (SED) that can convert this
bolometric luminosity into time-evolving spectra.  From this spectral
series, synthetic photometry in any desired filter and at any desired
redshift can be calculated for comparison to observed data.  We
discuss each of these ingredients in turn.

\subsection{Magnetar model}
\label{sec:Magnetar}

\begin{figure}
\includegraphics[scale=0.5]{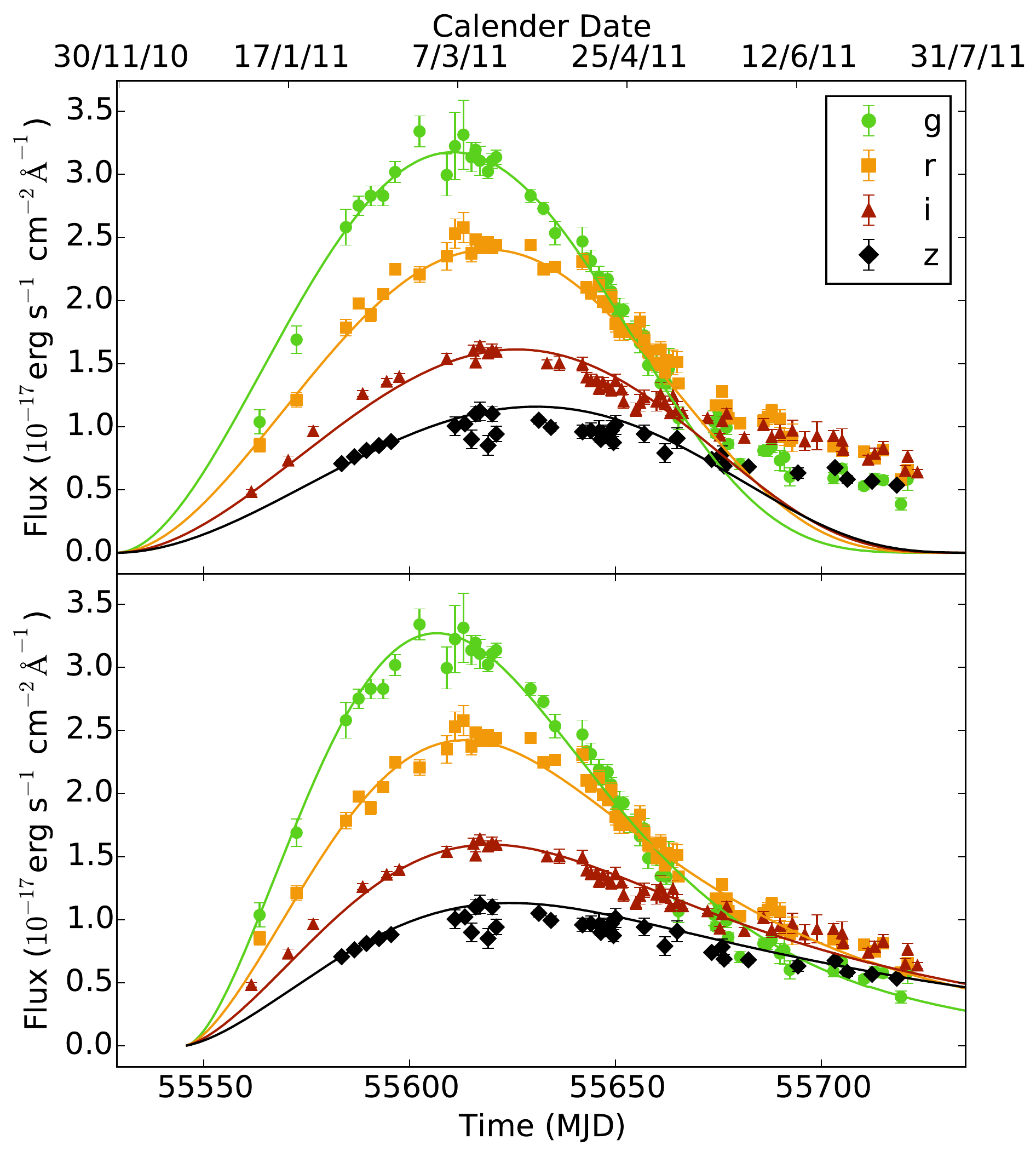}
\caption{The SLSN-I PS1-11ap $griz$ light curve
  \citep{2014MNRAS.437..656M} compared to two models describing its
  photometric evolution. In the upper panel, the model is a simple
  expanding and cooling black body fitted to data around maximum light
  only, and in the lower panel the model is our `absorbed' magnetar
  model fitted to the entire light curve. In the case of the magnetar model, the spectrum of
  SNLS-06D4eu \citep{2013ApJ...779...98H} has been used as an
  absorption template in the modelling of the SED (see Section
  \ref{sec:KCorrection}). Note that while both models can produce
  reasonable fits around the peak of the light cuve, the black body
  model is not able to reproduce the characteristic late-time
  behaviour of SLSNe.}
\label{fig:PS1-11ap}
\end{figure}

Early studies used a simple expanding photosphere radiating as a
cooling black body to represent the SLSN-I light curves \citep[eg.
][]{2013ApJ...779...98H}. This approach provides a good approximation
around the peak of the light curve, but it begins to significantly
deviate from the data at 30 days after maximum light
(Fig.~\ref{fig:PS1-11ap}).  Instead we use the currently popular
magnetar model, which is able to reproduce the photometric behaviour
for the entire SLSN-I population
\citep{2013ApJ...770..128I,2013Natur.502..346N}, in particular at late
times.

In the magnetar model, the bolometric luminosity, $L$, as a function
of time $t$ for a homologously expanding ejecta is
\citep{1982ApJ...253..785A}
\begin{equation}
L(t) = 4.9\times 10^{46}\,e^{ -(t / \tau_\mathrm{m})^2 }\delta(t) \int_{0}^{t} \frac{2t'}{\tau_\mathrm{m}^2}\,e^{(t'/\tau_\mathrm{m})^2}\,\frac{B_{14}^{2}\,P_{\mathrm{ms}}^{-4}}{\left(1+t'/\tau_\mathrm{p}\right)^2} dt',
\label{Eq:MagnetarLum}
\end{equation}
where $\tau_\mathrm{m}$ is the diffusion timescale, $B_{14}$ is the
neutron star magnetic field in units of $10^{14}$\,G,
$P_{\mathrm{ms}}$ is the magnetar spin period in ms, $\delta(t)$ is
the deposition function or trapping coefficient, and $\tau_\mathrm{p}$
is the magnetar spin-down timescale, inferred from $B_{14}$ and
$P_{\mathrm{ms}}$ \cite[see Appendix D of ][and references therein for
full details]{2013ApJ...770..128I}.

The trapping coefficient, $\delta(t)$, is often assumed to be unity
(i.e., implying the full trapping of the high-energy photons radiated
by the magnetar within the supernova ejecta) and time-independent
\citep{2013ApJ...770..128I,2015MNRAS.449.1215P,2015MNRAS.452.3869N}.
Here we use a correction introduced by \cite{2015ApJ...799..107W} with
a time-dependent trapping coefficient of
\begin{equation}
\delta(t) = 1 - \exp\left({-\frac{9\kappa \mathrm{M}_{\mathrm{ej}}^{2}}{40\pi  E_k} t^{-2}} \right),
\label{Eq:Wang}
\end{equation}
where $\mathrm{M}_{\mathrm{ej}}$ is the ejecta mass, $E_k$ is the
explosion energy, and $\kappa$ is the opacity.
$\mathrm{M}_{\mathrm{ej}}$ is proportional to $\tau_\mathrm{m}$, $E_k$
and $\kappa$ \citep{2013ApJ...770..128I}. We fix the explosion energy
to be $E_k = 10^{51}$\,erg and the opacity as $\kappa =
0.1$\,cm$^2$g$^{-1}$ following other studies
\citep[e.g.][]{2013ApJ...770..128I,2014ApJ...796...87I,2015MNRAS.452.3869N,2015MNRAS.449.1215P}.
Using this time-dependent trapping coefficient improves the late-time
fit to the light curve.  For a typical SLSN we calculate $\delta
\simeq 1$ up to 75 days post explosion. However, as the ejecta opacity
to high energy photons decreases with time we find $\delta \simeq 0.8$
at 150 days post explosion, emphasising the importance of this
correction in the late time light curves.

\pagebreak
We use the equations derived in Appendix D of
\cite{2013ApJ...770..128I} to relate the magnetar model parameters
described above to the photospheric radius and temperature and their
time evolution. The photospheric radius is proportional to the ejecta
velocity, which is also a function of the kinetic energy and
$\mathrm{M}_{\mathrm{ej}}$.  Thus, using Planck's law the magnetar
model parameters can also be used to generate a simple smooth SED. In
the next section, we discuss how we adjust this SED to physically
resemble the spectra of SLSNe-I.

\subsection{Spectral energy distributions}
\label{sec:KCorrection}

The spectra of SLSNe-I are relatively featureless in the rest-frame
optical, with characteristic broad lines of \ion{O}{ii}, and evolve slowly
\citep{2011ApJ...743..114C,2013ApJ...779...98H,2015MNRAS.449.1215P,2014ApJ...797...24V}.
However, there are much stronger absorption features in the rest-frame
ultraviolet (UV), with features attributed to \ion{Mg}{ii},
\ion{Fe}{iii}, \ion{C}{ii}, \ion{Co}{iii}, \ion{Si }{iii} and
\ion{Ti}{iii} \citep[see][for line
identifications]{2016MNRAS.458.3455M}. This UV SED is of prime
importance for our study, as it is redshifted into the optical at high
redshift where our search is most sensitive and where we probe the
largest volume. Thus it is important to construct our magnetar model
with an appropriate SED for our $k$-corrections to be realistic.

\begin{figure}
\includegraphics[scale=0.5]{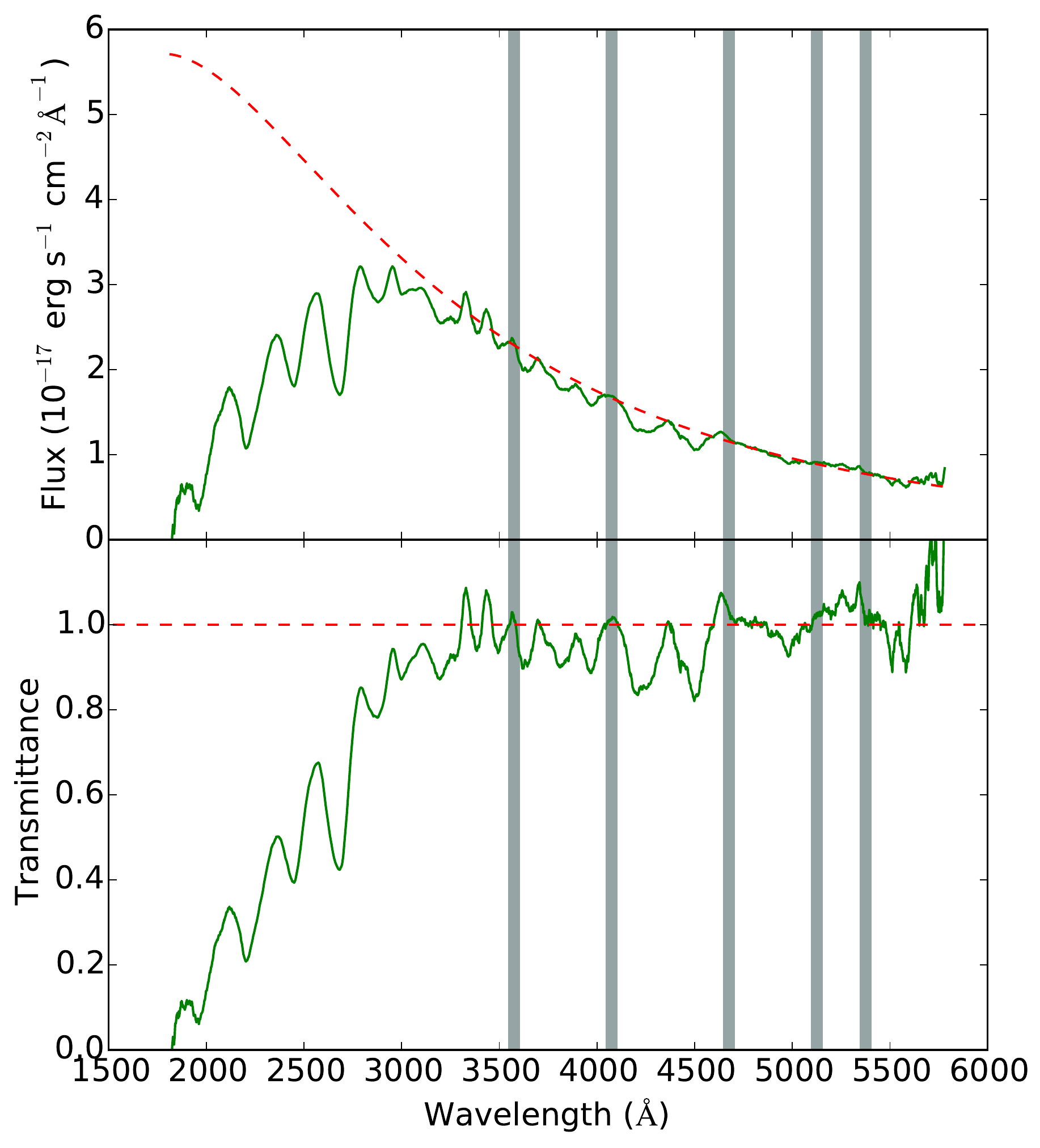}
\caption{Contructing template SEDs for SLSNe-I. Upper panel: the
  spectrum of iPTF13ajg \citep[solid;][]{2014ApJ...797...24V} fitted to
  Planck's law (dashed) in narrow, 50\,\AA, continuum regions (vertical
  bands).  There is a good agreement between the black body and
  the data in the region $\lambda>3000$\,\AA, with stronger absorption
  features appearing further blueward. Lower panel: the ratio between
  the observed spectrum and the continuum fit giving the absorption
  strength as a function of wavelength. This can then be used to
  improve the accuracy of the UV SED by combining it with a time
  dependent black body (Section \ref{sec:KCorrection}).}
\label{fig:specTemplate}
\end{figure}

The number of SLSNe-I with good UV coverage remains small. We
construct SED templates from three example SLSN which have a good
coverage in the UV: iPTF13ajg \citep{2014ApJ...797...24V}, SCP06F6
\citep{2009ApJ...690.1358B} and SNLS-06D4eu
\citep{2013ApJ...779...98H}. In each case we use one spectrum per
object, as spectral time series are only available for iPTF13ajg and SCP06F6; for
these objects we use the spectrum closest to maximum light. We follow
\cite{2014ApJ...797...24V}, fitting Planck's law to several
featureless, 50\,\AA\ wide, continuum regions in the observed spectra
of our three events with a good UV coverage (Fig.~\ref{fig:specTemplate}), in
order to estimate the black body continuum. We then use the ratio
between the observed spectra and these blackbody continua as measure
of the strength of the absorption features as a function of wavelength
in the different spectra. The result is a multiplicative function
that, when combined with a black body continuum from Planck's law, can
reproduce an observed SLSN-I spectrum on any epoch
(Fig.~\ref{fig:specTemplate}). This combination of the magnetar model
and our UV SED templates results in a significant improvement in the
light curve fits compared to the simpler expanding and cooling black
body model (Fig.~\ref{fig:PS1-11ap}).

\subsection{Superluminous supernova definition}
\label{sec:Definition}

Our final step is to select parameters of the magnetar model that
define the extent of SLSN-I parameter space using a training sample of
suitable events.  Only a small number of published spectroscopically
confirmed SLSNe-I have data of sufficient quality to constrain their
luminosity, rise and decline time, as well as the colour evolution. We
therefore introduce data-quality cuts in the published sample: we
select only objects observed with a minimum of two epochs in at least
three filters before maximum light, and two epochs in at least three
filters between maximum light and $+30$ days, where maximum light is
measured in the rest-frame $u$-band. 15 events which pass these cuts
and form our training sample are listed in
Table~\ref{tab:PubishedSLSNe}. We have corrected all the published
photometry for Milky Way extinction, and set an arbitrary error floor
of 0.03 magnitudes for all photometric points to account for possible
systematic uncertainties that were not included in the published data.

\begin{table*}
\begin{center}
  \caption{The training sample of SLSNe-I.}
\label{tab:PubishedSLSNe}
\begin{tabular}{|l|r|l|l|}
\hline
  \multicolumn{1}{|c|}{SN Name} &
  \multicolumn{1}{c|}{Redshift} &
  \multicolumn{1}{c|}{Survey} &
  \multicolumn{1}{c|}{Reference} \\
\hline
  PTF12dam & 0.108 & Palomar Transient Factory & \cite{2013Natur.502..346N}\\
  SN2011ke & 0.143 & Catalina Real-Time Transient Survey & \cite{2013ApJ...770..128I}\\
  &&\& Palomar Transient Factory \\
  SN2012il & 0.175 & Pan-STARRS & \cite{2013ApJ...771...97L}\\
  PTF11rks & 0.192 & Palomar Transient Factory & \cite{2013ApJ...770..128I}\\
  SN2010gx & 0.230 & Palomar Transient Factory & \cite{2010ApJ...724L..16P}\\
  PTF09cnd & 0.258 & Palomar Transient Factory & \cite{2011Natur.474..487Q} \\
  SN2013dg & 0.265 & Catalina Real-Time Transient Survey & \cite{2014MNRAS.444.2096N} \\
  PS1-11ap & 0.524 & Pan-STARRS & \cite{2014MNRAS.437..656M}\\
  DES14X3taz & 0.608 & Dark Energy Survey & \cite{2016ApJ...818L...8S} \\
  PS1-10bzj & 0.650 & Pan-STARRS & \cite{2013ApJ...771...97L}\\
  DES13S2cmm & 0.663 & Dark Energy Survey & \cite{2015MNRAS.449.1215P} \\
  iPTF13ajg & 0.741 & intermediate Palomar Transient Factory &\cite{2014ApJ...797...24V}\\
  PS1-10awh & 0.909 & Pan-STARRS & \cite{2011ApJ...743..114C}\\
  SNLS-07D2bv & 1.500 & SNLS &\cite{2013ApJ...779...98H}\\
  SNLS-06D4eu & 1.588 & SNLS &\cite{2013ApJ...779...98H}\\
\hline\end{tabular}
\end{center}
\end{table*}

We fit the training sample light curves with the magnetar model.  We
calculate the bolometric luminosity from eqn.~(\ref{Eq:MagnetarLum})
and eqn.~(\ref{Eq:Wang}), and estimate the photospheric radius using
equations from \citet{2013ApJ...770..128I}. The Stefan-Boltzmann law
then gives the photometric temperature. From these, we calculate the
black body SED from Planck's law, apply the absorption template, place
at the correct redshift, and integrate through the observed filters on
the epochs data were obtained. We use all three spectral absorption
templates for each event, and retain only the best fitting template
(see Table~\ref{table:Magnetar}).

The parameter estimation is performed using \textsc{MultiNest}
\citep{2008MNRAS.384..449F,2009MNRAS.398.1601F,2013arXiv1306.2144F},
an implementation of the nested sampling algorithm that is
particularly suited to complex probability distributions, run in
multi-modal mode. Our fit parameters are $\tau_\mathrm{m}$, $B_{14}$,
$P_{\mathrm{ms}}$, and the explosion date, $t_\mathrm{expl}$. We also
fit for host-galaxy extinction, parametrised using the colour excess
$E(B-V)_\mathrm{host}$, using the small magellanic cloud (SMC)
extinction law \citep{2003ApJ...594..279G} with $R_V=2.7$. This
extinction law is measured in environments that most likely resemble
the metal poor and star-forming dwarf galaxies associated with the
hosts of SLSNe-I. We use a uniform prior on all fit parameters with the
boundary values (Table \ref{tab:Prior}) chosen arbitrarily to allow a
large margin between the prior edge and the fit parameters for the
training sample of SLSNe-I. For host galaxy extinction, we allow
$E(B-V)_\mathrm{host}$ to vary between 0 and 0.2\,mag, typical for
extinctions measured directly from spectroscopy of SLSN-I host
galaxies \citep{2015MNRAS.449..917L}, and typical of extinctions
present in low stellar-mass galaxies \citep{2010MNRAS.409..421G}.

The magnetar model provides a good fit to all the SLSNe-I that we
selected for our sample. Table~\ref{table:Magnetar} contains the fit
parameters, as well as two additional derived parameters: the peak
absolute magnitude in the Sloan Digital Sky Survey $u$-band filter
\citep[$M_u$;][]{2000AJ....120.1579Y}, and the rise-time, $t_\mathrm{rise}$, measured from explosion
to maximum light in the rest-frame $u$-band. Note that the $M_u$ are
given in the AB magnitude system; $M_u^{\mathrm{AB}}\simeq
M_u^{\mathrm{vega}}+0.9$ \citep{2007AJ....133..734B}. Thus while some
of the training sample have $M_u^{\mathrm{AB}}>-21$, none have
$M_u^{\mathrm{vega}}>-21$.

\begin{table}
\begin{center}
  \caption{Priors on the fit parameters in the magnetar model}
\label{tab:Prior}
\begin{tabular}{|l|c|c|}
\hline
  \multicolumn{1}{|c|}{Parameter} &
  \multicolumn{1}{c|}{Lower limit} &
  \multicolumn{1}{c|}{Upper limit} \\
\hline
$\tau_\mathrm{m}$ (days) & 10 & 100 \\
$B_{14}$ ($10^{14}$\,G) & 0.1 & 20 \\
$P_{\mathrm{ms}}$ (ms) & 1 & 20 \\
$E(B-V)_\mathrm{host}$ & 0.001 & 0.2 \\
\hline\end{tabular}
\end{center}
\end{table}

\begin{figure*}
\includegraphics[scale=0.50]{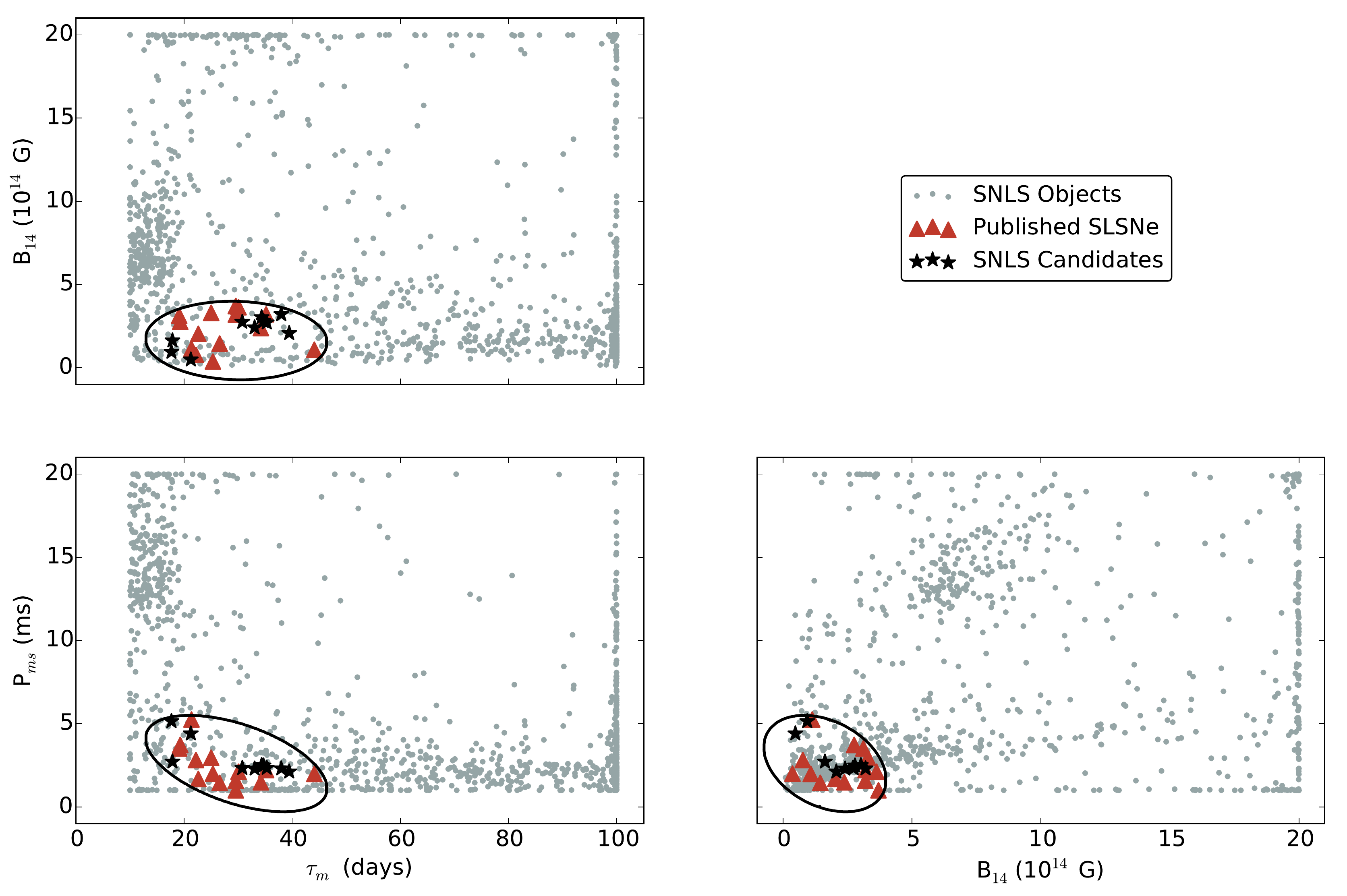}
\caption{The $\tau_\mathrm{m}$--$B_{14}$--$P_{\mathrm{ms}}$ parameter space
  constructed from the magnetar model fits as described in
  Section~\ref{sec:Magnetar}. The SNLS objects are denoted by grey
  circles. The ellipses correspond to the two dimensional
  projections of the three dimensional ellipsoid, fitted around the
  parameter space of the known SLSNe-I (shown as triangles) to
  form a region defining them in terms of the model. The SNLS
  candidates that fall within this parameter space are shown as
  stars.}
\label{fig:SLAPparam}
\end{figure*}

Fig.~\ref{fig:SLAPparam} shows the best fit magnetar model parameter
space. We define the SLSN-I parameter space using an ellipsoid that is
the lowest volume, simple geometric body consistent will all the
SLSN-I training sample.  We use the Khachiyan algorithm
\citep{Aspvall19801,KHACHIYAN198053} to find the smallest volume
enclosing all points (see Appendix~\ref{sec:slsn-parameter-space} for
the parametrisation).  Fig.~\ref{fig:SLAPparam} shows the three
two-dimensional projections of this parameter space, populated with
our training sample of literature SLSNe-I.

\begin{table*}
\begin{center}
\caption{Magnetar model parameters for the sample of 15 published SLSNe.}
\label{table:Magnetar}
\begin{tabular}{|l|r|r|r|r|r|r|r|r|r|r|}
\hline
  \multicolumn{1}{|c|}{Name} &
  \multicolumn{1}{c|}{$M_u$} &
  \multicolumn{1}{c|}{$t_\mathrm{rise}$} &
  \multicolumn{1}{c|}{$\tau_\mathrm{m}$} &
  \multicolumn{1}{c|}{$B_{14}$} &
  \multicolumn{1}{c|}{$P_{\mathrm{ms}}$} &
  \multicolumn{1}{c|}{$t_\mathrm{expl}$} &
  \multicolumn{1}{c|}{$E(B-V)$} &
  \multicolumn{1}{c|}{$\chi^2_{\nu}$} &
  \multicolumn{1}{c|}{Template} \\ & &
  \multicolumn{1}{c|}{(days)} &
  \multicolumn{1}{c|}{(days)} &
  \multicolumn{1}{c|}{($\times10^{14}$ G)} &
  \multicolumn{1}{c|}{(ms)} &
  \multicolumn{1}{c|}{(MJD)} & \\
\hline
  PTF12dam & -21.4 & 34.8 & 22.1 & 0.78 & 2.82 & 56044.8 & 0.17 & 2.47 & iPTF13ajg\\
  SN2011ke & -21.3 & 24.7 & 30.0 & 3.63 & 2.11 & 55647.5 & 0.019 & 1.27 & SNLS-06D4eu\\
  SN2012il & -21.2 & 21.2 & 19.1 & 3.12 & 3.52 & 55912.4 & 0.19 & 1.99 & SCP06F6\\
  SN2010gx & -21.7 & 24.3 & 29.6 & 3.19 & 1.57 & 55247.9 & 0.16 & 0.86 & SNLS-06D4eu\\
  PTF09cnd & -22.1 & 43.4 & 40.0 & 0.98 & 1.72 & 55024.3 & 0.02 & 0.32 & SNLS-06D4eu\\
  SN2013dg & -21.2 & 24.0 & 25.0 & 3.29 & 2.96 & 56415.1 & 0.15 & 0.14 & iPTF13ajg\\
  PTF09atu & -21.7 & 29.6 & 19.4 & 0.98 & 2.63 & 55009.2 & 0.037 & 0.63 & iPTF13ajg\\
  PS1-11ap & -21.8 & 45.9 & 44.1 & 1.08 & 1.99 & 55540.6 & 0.094 & 0.31 & iPTF13ajg\\
  DES14X3taz & -21.6 & 49.2 & 26.6 & 0.12 & 1.28 & 57018.1 & 0.061 & 1.45 & SNLS-06D4eu\\
  DES13S2cmm & -19.9 & 31.5 & 21.4 & 1.14 & 5.25 & 56510.0 & 0.04 & 0.46 & SCP06F6\\
  PS1-10bzj & -21.2 & 22.4 & 19.3 & 2.76 & 3.73 & 55523.9 & 0.15 & 0.37 & SNLS-06D4eu\\
  iPTF13ajg & -22.4 & 28.4 & 26.6 & 1.45 & 1.45 & 56353.8 & 0.11 & 0.22 & iPTF13ajg\\
  PS1-10awh & -21.9 & 28.7 & 34.2 & 2.38 & 1.48 & 55461.4 & 0.01 & 0.15 & iPTF13ajg\\
  SNLS-07D2bv & -21.1 & 28.3 & 35.2 & 3.21 & 2.22 & 54132.2 & 0.05 & 0.37 & iPTF13ajg\\
  SNLS-06D4eu & -21.9 & 21.5 & 29.6 & 3.70 & 1.00 & 53956.1 & 0.16 & 1.20 & SNLS-06D4eu\\
\hline\end{tabular}
\end{center}
\end{table*}

The SLSN-I parameter space can now be used to classify
further objects that are not in our training sample by fitting the
magnetar model to those data and examining where their resulting best fits
lie compared to the parameter space defined by the training
sample. Ideally, we would now test this by fitting a second sample of
known SLSN-I events. However, the number of events is so small that it
is not currently possible to construct such a sample, and have enough
objects to construct the parameter space in Fig.\ref{fig:SLAPparam}.
Thus we assume that the parameter space in
Fig.~\ref{fig:SLAPparam} is defined by a representative sample of
events. We also note that the fitting method and SLSN-I definition makes
no assumption about the luminosity of the event; it is quite possible
for fit events to have $M_u>-21$ and still be classified as SLSNe.

While the magnetar model is used to describe the physical processes
behind SLSNe-I, it is possible that, due to the flexibility of the
model, it could also produce a good fit to a SLSNe-II.  We are unable
to test this at present as there are currently no publicly available
light curves of SLSNe-II which pass our quality cuts. We therefore
refer to events that lie in our parameter space as `SLSN-I-like'
events. We thus calculate the rate of objects which are
well-represented by this model and are similar to the published SLSNe.
This does not exclude the possibility that there may be further
objects of this class that we are unable to identify in our sample.

\section{The supernova legacy survey dataset}
\label{sec:Selecting}

We next describe the observational dataset on which we will make the
volumetric rate measurement. We first introduce the survey itself, and
then discuss our selection of candidate SLSNe from the archive of
variable objects.

\subsection{The Supernova Legacy Survey}
\label{sec:SNLS}

The SLSN-I dataset for our volumetric rate calculation is taken from
the Supernova Legacy Survey
\citep[SNLS;][]{2006A&amp;A...447...31A,2010A&amp;A...523A...7G}. SNLS
was a rolling transient survey using the MegaCam detector
\citep{2003SPIE.4841...72B} at the 3.6-m Canada-France-Hawaii
Telescope, operating from 2003-2008. The survey imaged four,
one-square degree deep fields \citep[D1 to D4; field centres can be
found in][]{2006AJ....131..960S} in four SDSS-like filters
($g_M$,$r_M$,$i_M$,$z_M$) every 3--5 days during dark and grey time,
over six months in each year. The photometric search was accompanied
by a dedicated spectroscopic follow-up survey that classified
$\sim$500 supernovae
\citep{2005ApJ...634.1190H,2008ApJ...674...51E,2009A&amp;A...507...85B,2011MNRAS.410.1262W}.
Although the search continued throughout 2008, the original $i_M$-band
filter was damaged in July 2007 \citep[see][]{2013A&A...552A.124B} and
data from beyond this date is not included in this analysis. This
leaves just over four years of imaging used here.

The SNLS detected 4949 transient objects using a difference-imaging
pipeline \citep{2010AJ....140..518P}, including many objects visually
designated as active galactic nuclei (AGN) and variable stars, as well as supernovae.  The
pipeline used to detect the transients and perform photometry was
designed to operate in real-time on fast timescales, usually $<$24
hours, to allow for a rapid spectroscopic prioritisation
\citep{2006AJ....131..960S} and follow-up \citep{2010AJ....140..518P}.
We have since performed more accurate `offline' photometry for all the
SNLS candidates using a well-established supernova photometry pipeline
\citep[see][and references therein]{2015MNRAS.446.3895F}. In addition,
for our best candidates, we have measured multi-season light curves to
check for long-term activity typical of AGN.

\subsection{Identifying SLSNe in SNLS}
\label{sec:ID}

SNLS spectroscopically confirmed two SLSNe-I:
SNLS-07D2bv at $z=1.500$ and SNLS-06D4eu at $z=1.588$
\citep{2013ApJ...779...98H}. Both objects were initially classified
from real-time photometry \citep{2006AJ....131..960S} as low
probability SN Ia candidates, and targeted for low priority
spectroscopic follow-up. These objects form part of our training
sample of published SLSNe-I (Section \ref{sec:Definition}).

Two further SLSN-I candidates have since been detected in deep, stacked
SNLS images \citep{2012Natur.491..228C}, with host galaxy redshifts of
$z=2.04$ and $z=3.89$. These objects were not, however, discovered in
the real-time SNLS pipeline as they were fainter than the single-epoch
detection limits, and thus do not form part of the sample in this
paper.

We search in the SNLS transient database for additional SLSN
candidates that are well fit by our model from
Section~\ref{sec:Defining}. We assign redshift information as follows.
1694 have spectroscopic redshifts, either from spectra of the
transients
\citep{2009A&amp;A...507...85B,2008A&amp;A...477..717B,2008ApJ...674...51E,2005ApJ...634.1190H}
or of the host galaxy from redshift catalogues in the SNLS search
fields \citep[e.g.,][]{2007ApJS..172...70L,2013A&A...559A..14L,2013PASA...30....1L}. Where
a spectroscopic redshift is not available, we use photometric redshift
estimates from \citet{2006A&amp;A...457..841I}, which provides
photometric redshift information for galaxies in the SNLS fields at
$i_M<25$. SNLS transients are associated to potential host galaxies by
selecting the nearest galaxy within a radius of 1.5\arcsec.  This
provides photometric redshift information on a further 1527 events. For
these transients, we use a range of redshift values in the fits
spanning the photometric redshift uncertainties.

We are then left with 1728 candidates with no redshift information.
Inspecting the light curves showed that only 292 of these were likely
to be real objects with multiple detections in multiple bands. The
remainder are likely false detections that incidentally matched the
SNLS real-time detection criteria \citep{2010AJ....140..518P}.
For these last objects, we assign a broad range of redshifts (0.2 $\leq$ z $\leq$ 1.6 in steps of
$\Delta$z = 0.1) and treat then identically to objects with a known redshift.
While we might naively expect the number of
hostless supernovae to be higher, the depth of the SNLS deep stacks is
good enough to measure photometric redshifts for all but the very
faintest host galaxies.

We fit all the SNLS objects using the three available absorption
templates, retaining the parameters corresponding to the best fit
only. During the fitting, we apply the same quality cuts as for our
training sample: two detections in at least three filters between the
fit explosion date and maximum light, and a further two detections in
at least three filters between maximum light and $+30$ days. Here, we
consider a detection to be $\geq5$\,$\sigma$ in the real-time
photometry. Of the 4949 SNLS transients, 2097 pass these data quality
cuts, and the position of their best-fitting magnetar model parameters
can be see in Fig.~\ref{fig:SLAPparam}.

As would be expected, our magnetar model is not a good fit to the
majority of the SNLS objects. We remove the bulk of these poor fits
using a conservative cut at a $\chi^2$ per degree of freedom
($\chi^2_{\nu}$) of 20. Such a large $\chi^2_{\nu}$ cut is designed to
retain all SLSNe (see Table~\ref{table:Magnetar} for typical
$\chi^2_{\nu}$ for SLSNe), even those where a part of the light curve
may not be well-represented by our model, such as the pre-peak `bump'
observed in some SLSNe
\citep{2015ApJ...807L..18N,2015arXiv151103740N,2016ApJ...818L...8S}.

Of the SNLS objects that pass our data quality and $\chi^2_{\nu}$
cuts, 12 lie within the parameter space of SLSNe-I as defined in
Section~\ref{sec:Defining}. This includes the two spectroscopically
confirmed events from \citet{2013ApJ...779...98H} that were part of
our training sample.  Visual inspection of the light curves of the
other 10 candidates showed that many of these display features
indicative of AGN variability, e.g., a weak colour evolution and
multiple maxima in the light curves.

For our 10 candidates, we therefore measure the multi-season,
light curves (see Section \ref{sec:SNLS}) and use them to
search for signatures of any long-term variability or detections prior
to, or sufficiently after, the putative supernova event. This process of
visual inspection was very effective, eliminating all but a single,
new SLSN-I-like candidate: SNLS-07D3bs. All of the light curves of the
9 candidates that passed the data quality and $\chi^2_{\nu}$ cuts and
that lie in the parameter space of SLSNe-I, but which do not pass
visual inspection, are shown for reference in
Appendix~\ref{sec:light-curves}.

\subsection{SNLS-07D3bs}
\label{sec:07D3bs}

\begin{figure}
\centering
\includegraphics[scale=0.5]{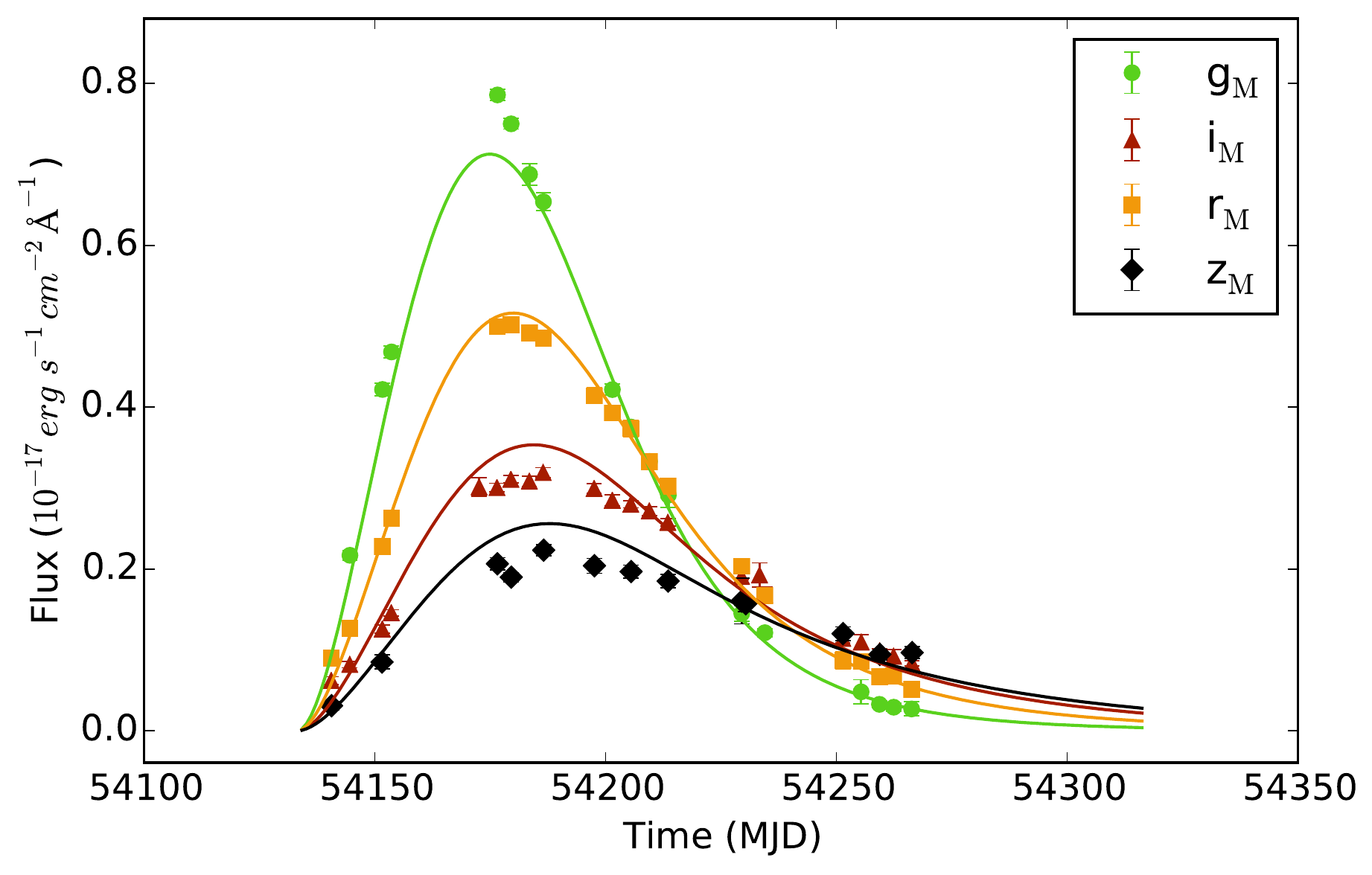}
\caption{The $g_M$, $r_M$, $i_M$, $z_M$ light curve of SNLS-07D3bs
  overplotted with the best fit magnetar model (Section
  \ref{sec:Defining}) at $z=0.757$. The candidate shows a good
  agreement with the model.}
\label{fig:Lightcurves}
\end{figure}

SNLS-07D3bs was identified as a SLSN candidate between $0.6<z<1.2$
based on its host galaxy photometric redshift estimate, with a best fit
to the magnetar model at $z\simeq1.0$ (see Fig.~\ref{fig:Lightcurves})
using the spectrum of SNLS-06D4eu as the UV template.  SNLS-07D3bs was
observed spectroscopically during SNLS on 2007 April 17 at the Keck-I
10-m telescope using the Low Resolution Imaging Spectrograph (LRIS),
but no spectral classification could be made from the data at the
time \citep{2013PhDT.......150F}.

\begin{figure}
\centering
\includegraphics[scale=0.5]{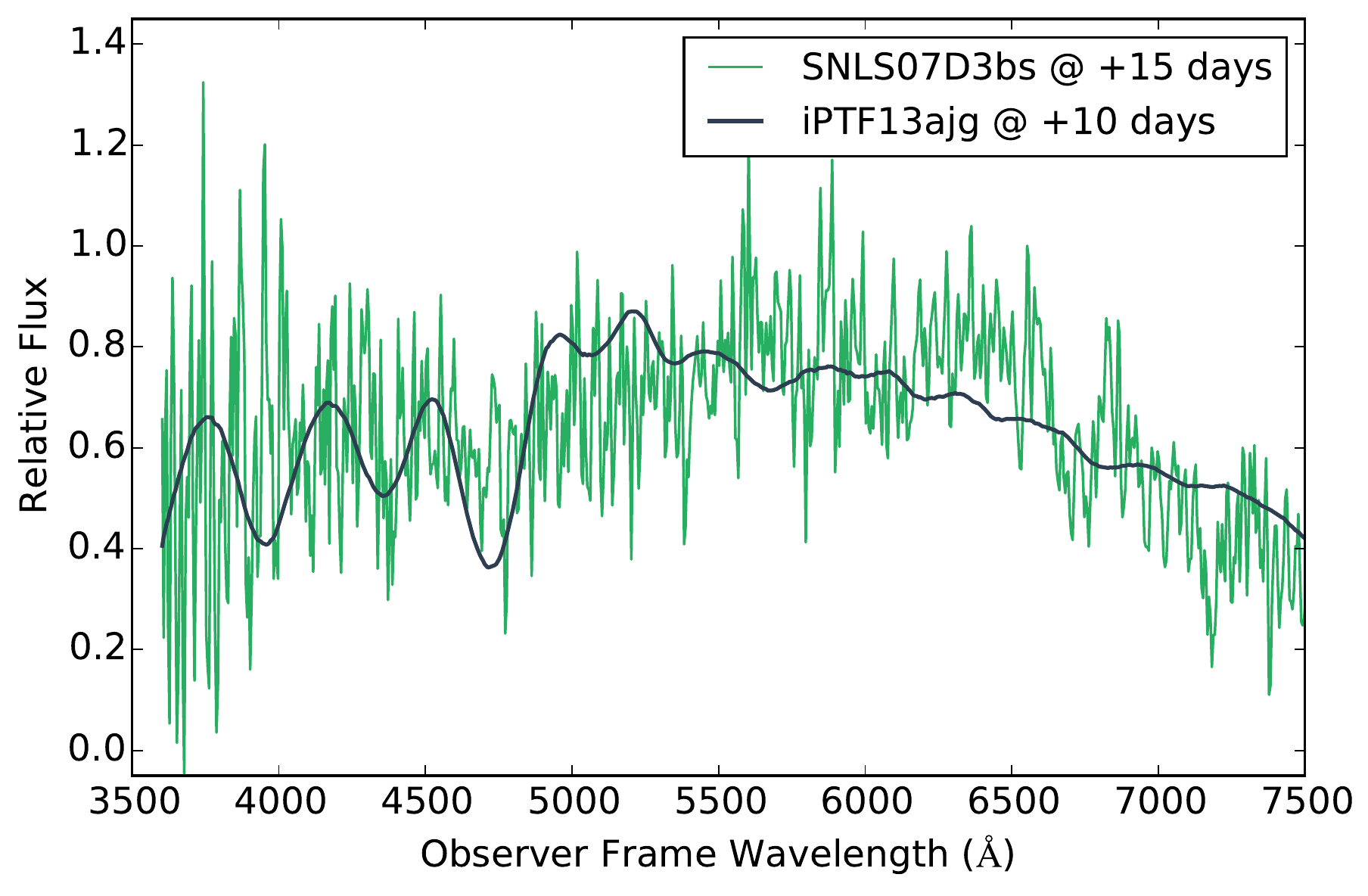}
\caption{The spectrum of SNLS-07D3bs from Keck-I/LRIS, taken 15
  rest-frame days after maximum light. The signal-to-noise of the
  spectrum is low preventing a definitive classification; however, the
  spectrum is consistent with a SLSN at around $z=0.76$. Weak galaxy
  emission lines are consistent with $z=0.757$.}
\label{fig:07D3bsSpec}
\end{figure}

However, at that time the SLSN-I class was not known and no SLSN-I
spectral templates were available for comparison with the data.
Therefore, using the \textsc{superfit} spectrum identification tool
\citep{2005ApJ...634.1190H}, we have re-analysed the spectrum of
SNLS-07D3bs. The data are noisy as observing conditions were quite
poor, but we find the best spectral match to be to the SLSN-I
iPTF13ajg at $z=0.757$ (Fig.~\ref{fig:07D3bsSpec}). While this is
clearly an uncertain spectral classification, there is no evidence
from the spectrum that the object is not a SLSN, and the best match is
an event of that type.  The magnetar model also provides a good fit at
that redshift (see Table~\ref{tab:snls07d3bs_params} for the magnetar
model fit parameters). The host galaxy (RA=14\fh21\fm50\fs466,
Dec.=+53\fd10\fm28\fs58) galaxy was detected in the SNLS deep stack
images \citep{2006A&amp;A...457..841I}, but is very faint, with AB
magnitudes of ($u_M$, $g_M$, $r_M$, $i_M$, $z_M$) = ($26.61\pm0.49$,
$26.13\pm0.16$, $25.67\pm0.13$, $25.18\pm0.11$, $25.19\pm0.37$).
Taking the evidence together, we consider SNLS-07D3bs to be the third
SLSN detected by the real-time pipeline of SNLS.

\begin{table*}
\begin{center}
\caption{Magnetar model parameters for the new SNLS SLSN candidate: SNLS-07D3bs.}
\label{tab:snls07d3bs_params}
\begin{tabular}{|l|r|r|r|r|r|r|r|r|r|r|}
\hline
  \multicolumn{1}{|c|}{Name} &
  \multicolumn{1}{c|}{$M_u$} &
  \multicolumn{1}{c|}{$t_\mathrm{rise}$} &
  \multicolumn{1}{c|}{$\tau_\mathrm{m}$} &
  \multicolumn{1}{c|}{$B_{14}$} &
  \multicolumn{1}{c|}{$P_{\mathrm{ms}}$} &
  \multicolumn{1}{c|}{$t_\mathrm{expl}$} &
  \multicolumn{1}{c|}{$E(B-V)$} &
  \multicolumn{1}{c|}{$\chi^2_{\nu}$} &
  \multicolumn{1}{c|}{Template} \\ & &
  \multicolumn{1}{c|}{(days)} &
  \multicolumn{1}{c|}{(days)} &
  \multicolumn{1}{c|}{($\times10^{14}$ G)} &
  \multicolumn{1}{c|}{(ms)} &
  \multicolumn{1}{c|}{(MJD)} & \\
\hline
SNLS-07D3bs & -20.9 &  27.2 & 23.7 & 2.28 & 3.81 & 54132.5 & 0.05 & 1.96 & iPTF13ajg\\
\hline
\end{tabular}
\end{center}
\end{table*}

\section{The rate of superluminous supernovae}
\label{sec:MC}

Having identified a sample of three SLSNe in SNLS, in this section we
now calculate the volumetric rate of SLSNe ($\rho_{\mathrm{slsn}}$)
implied by these detections.

\subsection{Method}
\label{sec:method}

$\rho_{\mathrm{slsn}}$ is defined as a sum over the $N$ SLSNe found in a
given comoving search volume $V$ over a search duration $T$, weighted
by the inverse of the detection efficiency, $\epsilon_{i}$, of
detecting each event, i.e.,
\begin{equation}
\label{eq:rate}
\rho_{\mathrm{slsn}} = \frac{1}{V}\sum^{N}_{i}\frac{(1+z_i)}{\epsilon_{i}T_{i}}.
\end{equation}
The factor $(1+z_i)$ corrects for time dilation. The detection
efficiency $\epsilon_i$ is a statistic describing how each SLSN should be
weighted relative to the whole population; $1-\epsilon_i$ gives the
fraction of similar SLSNe that exploded during the search period but
were missed by the survey due to (for example) search inefficiencies.

Our SLSN-I detection efficiencies are based on the analysis of
\citet{2010AJ....140..518P}, who carried out a study of the detection
efficiencies and selection biases of type Ia supernovae (SNe Ia) in
SNLS, and subsequently calculated a rate of those events in
\cite{2012AJ....144...59P}.  In this study, several million fake SNe
Ia were placed in the SNLS science images, with the correct temporal
evolution, and passed though the SNLS real-time detection pipeline.
The recovery efficiencies of these SNe Ia could then be estimated.
Although these results were produced using a particular model of a
particular supernova type, at a more basic level they also provide the
recovery efficiencies of point sources in the SNLS data as a function
of magnitude in \textit{every} $i_M$-band image that SNLS took, and it
is these more basic data that we use in this study.

We reverse the common approach to supernova rate calculations: instead
of calculating the rate of SLSNe starting with the number of detected
objects, we instead calculate the probability that a given input value
of $\rho_{\mathrm{slsn}}$ will lead to an eventual detection of three
SLSNe in the SNLS survey. This method also produces a non-Gaussian
uncertainly estimate as a by-product of the calculation in the form of
a rate probability distribution (Fig.~\ref{fig:rateFlat3}). We quote
the uncertainties as the 1-$\sigma$ confidence region of this
distribution.

We simulate the SLSN population using a Monte Carlo approach,
exploding SLSNe randomly within the SNLS search period and search
volume, and creating artificial SLSN light curves on each epoch on
which SNLS took data, including the effect of $1+z$ time dilation.
This gives an $i_M$ apparent magnitude on each epoch, which can be
compared to the point-source recovery statistics on that epoch to give
the probability of detection.

\subsection{Search volumes}
\label{sec:search-volumes}

The effective SNLS search areas in each field from which the search
volumes can be calculated can be found in \citep{2012AJ....144...59P};
the total search area is 3.56\,deg$^2$. The volume is calculated by
considering the redshift range to which our search is sensitive.
There is a small variation in the detection efficiency amongst the fields; D3 observing season was longer in comparison to the other fields while D1 and D2 had on average, marginally deeper exposures.

At the low-redshift end, the search volume is set by the redshift at
which a SLSN would become saturated in the SNLS data. At the
high-redshift end, the volume is set by the redshift at which we are
no longer able to recover a SLSN event, i.e. a SLSN would fall below
the detection limit of the survey.

\begin{figure}
\includegraphics[scale=0.5]{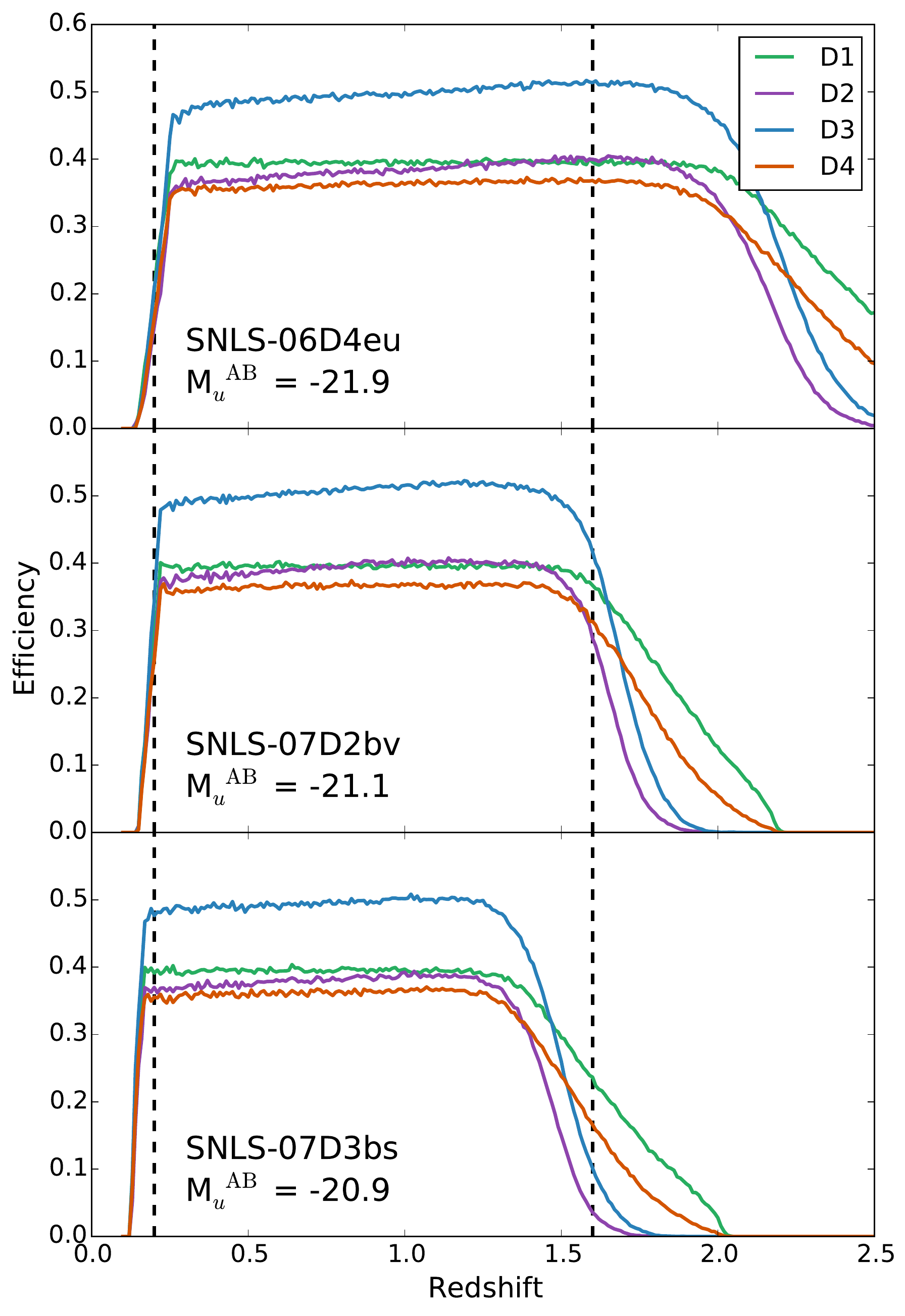}
\caption{The redshift range to which our SLSN search is sensitive, as
  a function of the four SNLS search fields.  The figure shows the
  recovery efficiency of three different SLSNe as a function of
  redshift, with each line corresponding to a different search field..
  The efficiency includes the same data quality cuts as used in the
  training sample in Section~\ref{sec:Definition} and the SNLS
  candidate selection in Section~\ref{sec:ID}. The vertical dashed
  lines at $z=0.2$ and $z=1.6$ illustrate the final redshift range
  used in our Monte Carlo rate calculations.}
\label{fig:zrange}
\end{figure}

Fig~\ref{fig:zrange} illustrates the redshift range to which we are
sensitive, showing the recovery efficiency as a function of redshift
for the three SNLS events from Section~\ref{sec:ID}. For events at
$z<0.2$, the efficiency drops rapidly due to saturation effects, and
thus we choose $z=0.2$ as the lower redshift limit. At the upper
redshift limit, we choose $z=1.6$; although events similar to
SNLS-06D4eu are detectable to beyond $z=2$, the fainter events like
SNLS-07D3ds will become undetectable in some of the SNLS search fields
beyond $z=1.6$.

\subsection{Rate calculation}
\label{sec:rate-calculation}

We begin each Monte Carlo simulation with an input
$\rho_{\mathrm{slsn}}$ value. From this we calculate the number of
SLSNe that would have occurred within the SNLS search area over the
redshift range to which we are sensitive ($0.2<z<1.6$) in bins of
$\Delta z = 0.01$. We assume that this rate does not evolve within the
redshift range which we consider in our simulation (we test this
assumption below). The artificial SLSNe are then assigned a random
spatial position (and consequent Milky Way extinction), redshift, host
galaxy extinction and physical parameters drawn from the magnetar
model parameter space derived from our training sample
(Fig.~\ref{fig:SLAPparam}). A random explosion epoch during the SNLS
search period is assigned to each event, and the predicted photometry
calculated on every epoch corresponding to a SNLS observation.

Using the point-source detection efficiencies of
\cite{2010AJ....140..518P}, we can then calculate the probability of
detecting each SLSN on every epoch of $i_M$ data, and combine the
probabilities to give the total probability of discovering each
object. In order to be considered detected, we also enforce that each
artificial SLSN must pass the same data quality cuts as both our
training sample (Section~\ref{sec:Definition}) and our real sample of
SNLS candidates (Section~\ref{sec:ID}). We repeat the entire
simulation 100,000 times over an input $\rho_{\mathrm{slsn}}$ range of $5
\leq \rho_{\mathrm{slsn}} \leq 500$\,SNe\,Yr$^{-1}$\,Gpc$^{-3}$.

\begin{figure}
\includegraphics[scale=0.5]{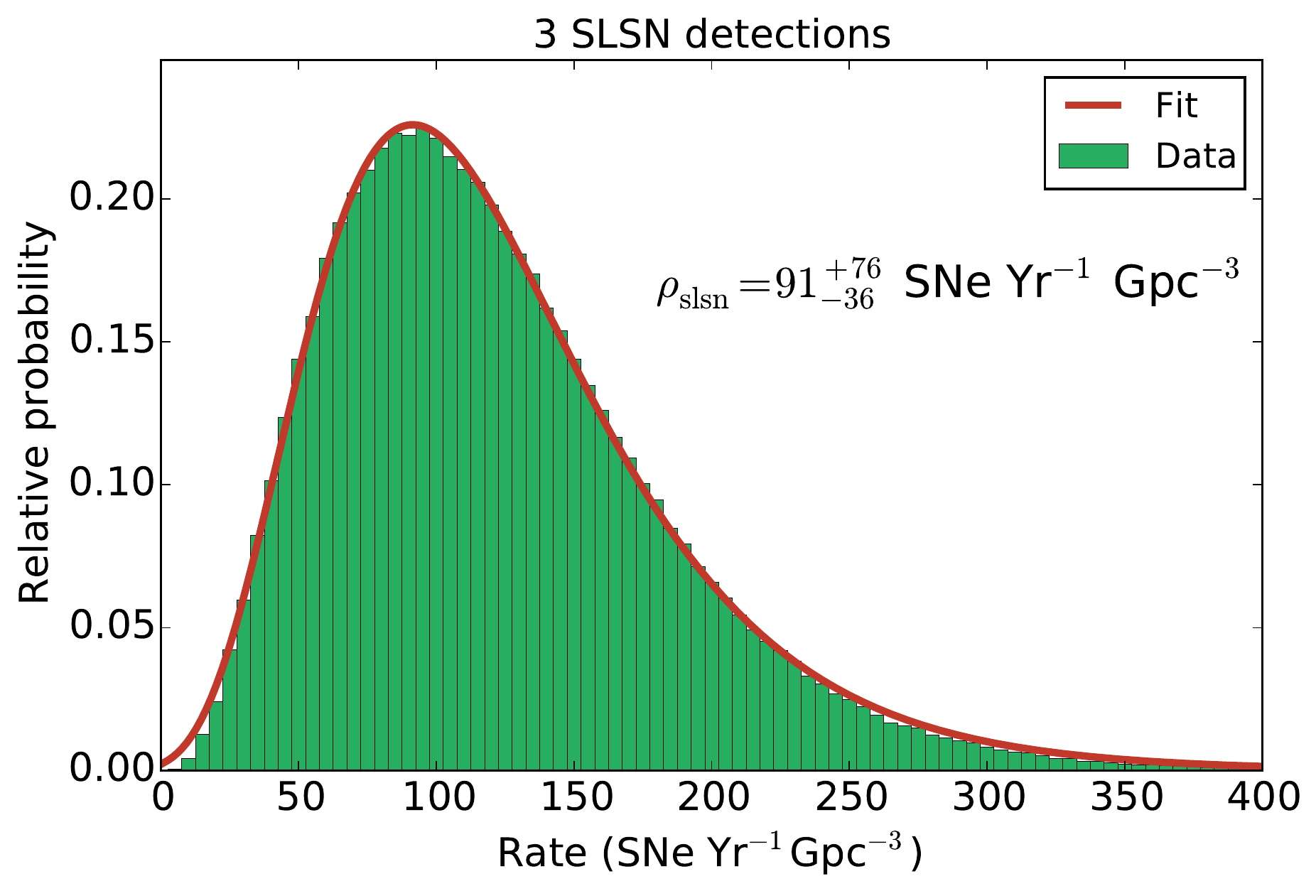}
\caption{The probability distribution of the volumetric rate of SLSNe
  for the three SLSN candidates over the duration of SNLS at $0.2<z <
  1.6$, as determined by our 100,000 Monte Carlo simulations. A
  log-normal distribution is fitted to the data (red line) to estimate
  the peak of the probability distribution and the uncertainties,
  quoted as the 68\% confidence region.}
\label{fig:rateFlat3}
\end{figure}

Fig.~\ref{fig:rateFlat3} shows the probability of three SLSN events
being detected in our simulations as a function of this input rate. A
log-normal distribution was fitted to the simulation results and used
to smoothly determine the peak of the distribution as well as the
$1\sigma$ confidence regions. From this, we find the rate of SLSNe at
$z=1.13$ (the volume-weighted centre of the $0.2 < z < 1.6$ range) to
be $\rho_{\mathrm{slsn}} =
91^{+76}_{-36}$\,SNe\,Yr$^{-1}$\,Gpc$^{-3}$.

We also investigate the effect that our assumption of a uniform
redshift distribution in the simulated SLSNe may have on our final
rate. We repeat the Monte Carlo simulation, but instead draw the SLSNe
from a redshift distribution that follows the cosmic star-formation
history \citep[SFH; taken from][]{2006ApJ...651..142H}. We find
$\rho_{\mathrm{slsn}} = 98^{+82}_{-39}$\,SNe\,Yr$^{-1}$\,Gpc$^{-3}$,
close to our original result and, considering the uncertainities, a
negligible difference.  Thus our final rate, averaged over the
redshift range we have considered, is not sensitive to the assumed
rate evolution in our simulation.  This is likely due to the relative
uniformity of our detection efficiency as a function of redshift
within our search volume (see Fig.~\ref{fig:zrange}).

\section{Discussion}
\label{sec:Discussion}
\begin{figure}
\includegraphics[scale=0.5]{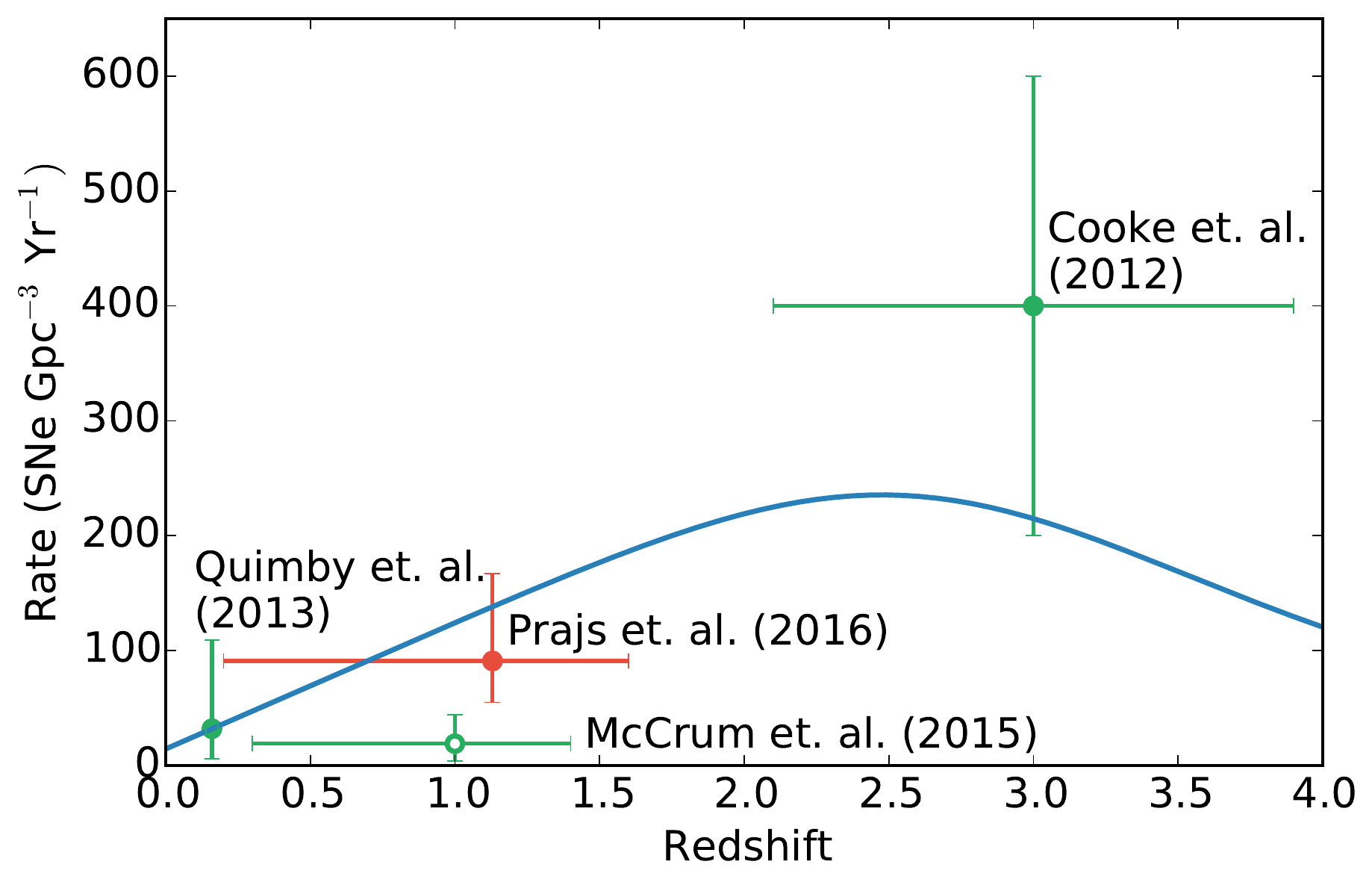}
\caption{The evolution of the volumetric SLSN rate as a function of
  redshift. We show measurements by \citet{2013MNRAS.431..912Q},
  \citet{2015MNRAS.448.1206M} and \citet{2012Natur.491..228C} for
  comparison. The \citet{2015MNRAS.448.1206M} result is marked by an
  open circle to highlight that it may not be directly comparable with
  the other measurements as it is derived by a comparison to the rate
  of core collapse supernovae and is not a direct measurement.  The
  observed evolution is consistent with that of the SFH over the same redshift range; we over-plot
  in blue the parametrisation of the cosmic SFH of
  \citet{2006ApJ...651..142H}, normalised to the low-redshift SLSN-I
  rate obtained by \citet{2013MNRAS.431..912Q}.}
\label{fig:rate}
\end{figure}

In Fig.~\ref{fig:rate}, we compare our new SLSN rate measurement with
other published values taken from the literature
\citep{2013MNRAS.431..912Q,2012Natur.491..228C} as a function of
redshift. We observe an increase in the volumetric rate as a function
of redshift. The extent of this observed evolution is consistent with
the evolution in the cosmic star-formation history (SFH) observed over
the same redshift range \citep{2006ApJ...651..142H}. This is, perhaps,
an unsurprising result, as SLSNe are thought to originate from the
death of very massive and hence short-lived stars
\citep{2015MNRAS.452.3869N,2015ApJ...807L..18N}. However, we note that
we cannot discriminate between the evolution that follows the SFH, and
one with the same evolution to $z=1.5$ followed by a peak at a much
higher redshift, as the $z>1.5$ measurement is quite uncertain.

A higher-redshift peak may be expected due to the association of SLSNe
with metal-poor galaxies. SLSNe-I almost invariably explode in
galaxies that are low-mass, compact dwarfs
\citep{2011ApJ...727...15N,2015ApJ...804...90L}, and that are
metal-poor and strongly star-forming
\citep{2013ApJ...771...97L,2013ApJ...763L..28C,2015MNRAS.449..917L}.
One popular interpretation of this is that the low metallicity must
play a role in the formation of SLSNe-I, which is consistent with the
low metal content inferred from their UV spectra
\citep{2016MNRAS.458.3455M}. This scenario would also predict a
volumetric rate evolution that follows both the cosmic SFH and cosmic
chemical enrichment. Further $z>2$ rate measurements are needed to
test this in more detail.

\begin{figure}
\includegraphics[scale=0.5]{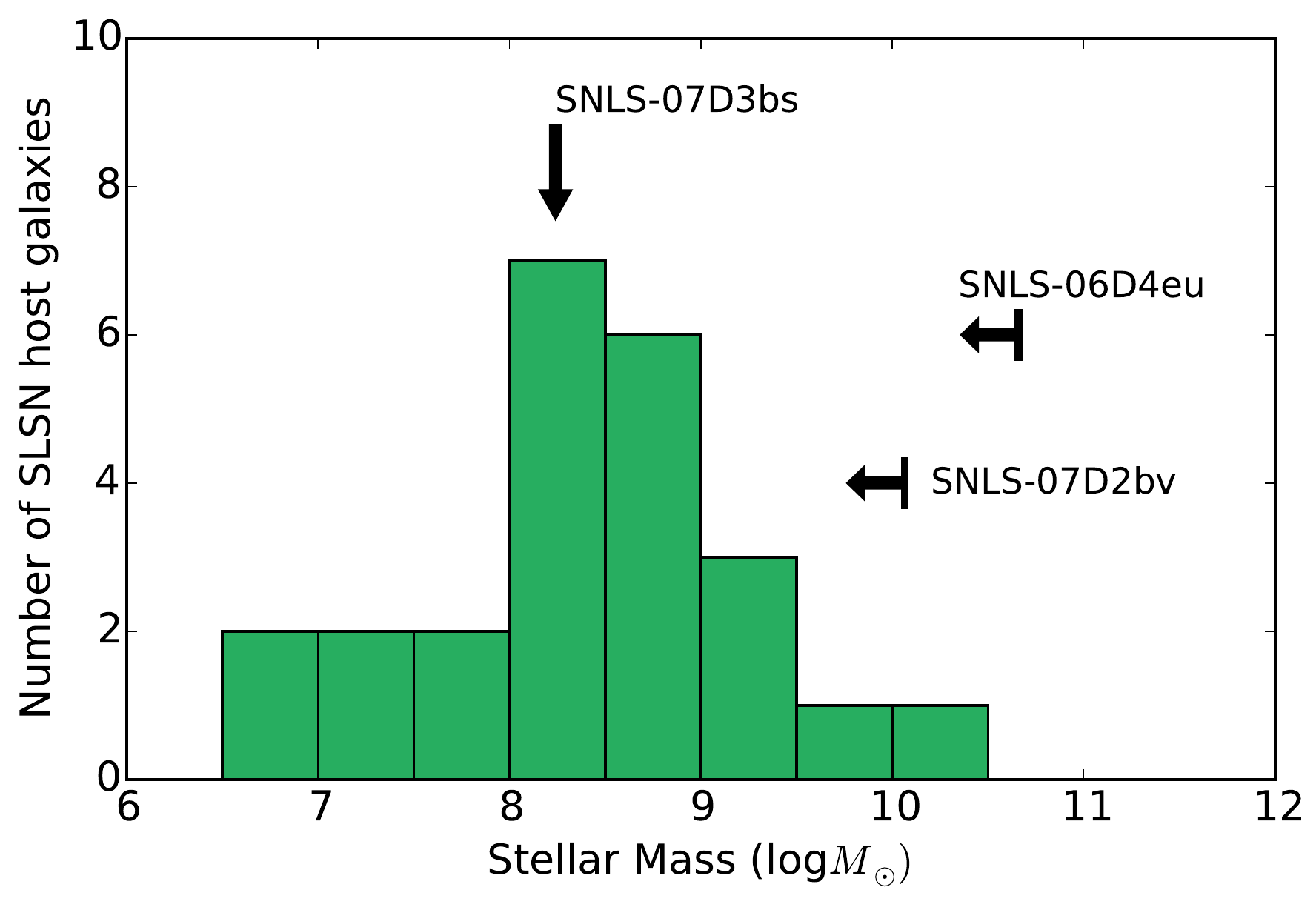}
\caption{The stellar mass distribution of SLSN host galaxies plotted using the data from \citet{2014ApJ...787..138L}, showing the consistency of SNLS07D3bs with the rest of the population. The lack of detections for the hosts of the high redshift candidates is consistent with being associated with low mass galaxies, found below the detection limit of SNLS at their redshifts.}
\label{fig:hosts}
\end{figure}

Fig.~\ref{fig:hosts} shows the distribution of SLSN host-galaxy
stellar masses as measured by \cite{2014ApJ...787..138L}. We use the
\textsc{zpeg} photometric redshift package
\citep{2002A&amp;A...386..446L} with the SNLS multi-waveband
$g_M$,$r_M$,$i_M$,$z_M$ host galaxy photometry to estimate the stellar
mass for SNLS-07D3bs. We do not attempt to derive host galaxy stellar
masses for SNLS-06D4eu and SNLS-07D2bv due to their faintness and the
lack of infrared (rest-frame optical) data.  Instead we place
conservative stellar mass limits, again derived using \textsc{zpeg}.
The host stellar masses and limits for all three of our candidates
agree with the published SLSN-I host stellar mass distribution
(Fig.~\ref{fig:hosts}).

Using SLSNe discovered by the Pan-STARRS medium deep survey over $0.3
\leq z \leq 1.4$, \citet{2015MNRAS.448.1206M} measure the relative
rate of SLSNe to be between $3^{+3}_{-2}\times10^{-5}$ and
$8^{+2}_{-1}\times10^{-5}$ that of the core-collapse supernova rate
($\rho_{\mathrm{cc}}$). We use the SNLS $\rho_{\mathrm{cc}}$ measurement
at $z\sim0.3$ of $1.42\pm0.6\times10^{-4}$\,SNe\,Gpc$^{-3}$\,yr$^{-1}$
 \citep{2009A&A...499..653B}, and extrapolate it to
$z=1.13$, assuming it tracks the cosmic SFH, increasing the rate by a factor of 2.62. Our own absolute SLSN
rate can then be expressed as rate relative to that of core collapse
SNe, which we find to be 2.2$^{+1.8}_{-0.9}\times10^{-4}$ of the
$\rho_{\mathrm{cc}}$. This is higher than, but still consistent with, the
relative rate of \citet{2015MNRAS.448.1206M}.

\subsection{Comparison with the rate of ULGRBs}

There have been recent suggestions \citep{2015Natur.523..189G} that
ultra-long gamma ray bursts \citep[ULGRBs;
e.g.,][]{2013ApJ...766...30G,2014ApJ...781...13L}, with prompt
gamma-ray emission lasting 10$^3$ to 10$^4$\,s, could also be powered
by the spin-down of a magnetar, in a similar mechanism to that
suggested for SLSNe, albeit with different initial conditions
explaining the physical differences between the phenomena
\citep{2014MNRAS.443...67M}. If both classes of objects were to
originate in a similar physical scenario, one might expect their
progenitors (and hence their rates) to be related.

The rate of ULGRBs is challenging to constrain. Bursts can be detected
by the \textit{Swift} satellite Burst Alert Telescope (BAT), but can
have different triggering criteria: objects can be detected based on
the detector rate over some given time period, or a search of the
image plane, or their total fluence. A burst that misses one trigger,
may still satisfy a different one at later times. This is particularly
true for very long events, where a burst may go undetected by the rate
triggers, or even by searches of the image plane on short time scales
(e.g., 64\,s). However they may then be recovered in image searches on
longer time-scales of $>1000$\,s, as is the case for several very long
events such as GRB\,101225A \citep{2011Natur.480...72T} and
\textit{Swift} J1644+57
\citep{2011Sci...333..199L,2011Natur.476..421B}.

However, it is still possible to crudely estimate upper and lower
limits to the rate of ULGRBs. The observed rate of a given individual
burst is given by \citep{2012MNRAS.425.2668C,2013ApJ...766...30G}
\begin{equation}
\rho_{\mathrm{ulgrb}} = \frac{1}{V_{\mathrm{max}} T \Omega \eta_z} f_b,
\end{equation}
where $V_{\mathrm{max}}$ is the maximum volume over which a burst
could be detected, $T$ is the time period over which the search was
detected (10 years for \textit{Swift}), the factor of $\Omega$
corrects for the fraction of the sky observed by \textit{Swift} at a
given time ($\Omega = 0.17$) and $\eta_z$ reflects that many bursts do
not have a known redshift, and so the true rate is higher by a
corresponding factor (we assume $\eta_z =0.3$). Finally, $f_b$ is the
beaming factor of the burst, reflecting the number of bursts that are
viewed off-axis to us and are hence undetected; $f_b = 2/\theta_j^2$,
where $\theta_j$ is the opening angle of the GRB jet.

Determination of $V_{\mathrm{max}}$ and $f_b$ can be difficult. In
particular, $V_{\mathrm{max}}$ depends on the evolution of the burst
properties (light curve and spectrum) with redshift, and how this is
then convolved with the instrument triggering thresholds. However, we
estimate this by assuming that the signal-to-noise of the detection
scales with the square of the distance, and hence calculate the
redshift at which a burst would pass the detection threshold
($7\sigma$). $f_b$ can be determined for a given burst by the
so-called `jet-break', at which the later expansion of the jet creates an
achromatic steepening of the observed afterglow emission.

We determine a pessimistic and realistic rate of ULGRBs. For our
pessimistic rate we include only the few well-studied ULGRBs (GRB
101225A, 111209A, 121027, 130925A) and omit the very long events
suggested to be tidal disruption flares. The maximum redshifts at
which these events could be detected are $z=0.9,1.3,2.3,0.6$
respectively, with the mean $V/V_{\mathrm{max}}=0.5$. As the mean
redshift of this sample is similar to that of our SLSN-I sample, we can
directly compared these measurements.  The rate of ULGRBs is then
$\rho_{\mathrm{ulgrb}} =7 \times 10^{-2} f_b$\,Gpc$^{-3}$\,yr$^{-1}$.

This is a however a lower limit, as it is clear from observations of
\textit{Swift} bursts that other events have similar properties, but
may not have been studied in depth at late times. The analysis of
\cite{2014ApJ...787...66Z} suggests that $\sim15$ bursts have engines
active for $>5000$\,s. Our estimated rate should therefore be increased
by a factor of four, assuming that these bursts belong to the same
class of events as ULGRBs.  This is however only a lower limit in the
correction as their analysis was unable to firmly locate the end of
the engine activity for $<50$ per cent of these events.

Additional objects could be added in the very longest outbursts,
typically assumed to be relativistic tidal flares
\citep{2011Sci...333..203B,2012ApJ...753...77C,2015MNRAS.452.4297B},
but with some similarities to SLSNe \citep{2016ApJ...819...51L}. The
longest bursts are difficult to detect because of the trigger
thresholds. Indeed, the very longest image triggers can only be used
in cases where the dwell of \textit{Swift} (i.e., how long it spends
in a given observation) is longer than the length of the image
trigger. This is only true for 15 per cent of the mission
\citep{2014ApJ...781...13L}, causing a further factor of six increase
in the rate that may be applicable for the longest events.
In practice this would increase the sample size modestly (by a factor
of two in extreme case). It should also be noted that the very long
events by no means uniquely trigger these very long triggers.
Therefore we estimate the true rate of ULGRBs to be
$\rho_{\mathrm{ulgrb}} \approx 0.1-0.6 f_b$\,Gpc$^{-3}$\,yr$^{-1}$,
with the upper limit comparable to, although slightly lower than, the
rate of long GRBs.

The final correction arises from the beaming rate. To date, jet-breaks
have not been observed in ULGRBs, and the limits on the beaming angles
(for typical ISM parameters) are around 12 degrees, indicating a
beaming correction factor of order 50. The upper range of the ULGRB
rate would then be $\rho_{\mathrm{ulgrb}} \approx
30$\,Gpc$^{-3}$\,yr$^{-1}$. This is interestingly close to the SLSN-I
rate we have measured at a similar redshift. If the physical mechanism
responsible for driving the SN in GRB 111209A is the same as
that responsible for other ULGRBs, and for creating the luminosity in
SLSNe-I, then one should expect that a ULGRB would be observed for
some observer in a significant fraction (and potentially all) of the
SLSNe-I.  This means that late-time observations could reveal off-axis
afterglow-like emission at radio or X-ray wavelength in SLSNe-I. While
one plausible example has been found in the bright late-time (100+
day) X-ray emission of SCP06F6 \citep{2013ApJ...771..136L}, other
observations may suggest that such a ratio is unlikely. In any case,
further multi-wavelength, late-time observations of SLSNe are clearly
motivated.

\section{Summary}
\label{sec:Summary}

In this paper, we have used data from the SNLS to calculate the volumetric rate of SLSNe
at $z\sim1.1$. We used a simple magnetar model, in conjunction
with new spectral templates, to develop a method for photometric
classification of SLSNe. After fitting the magnetar model to a set of
15 well-sampled SLSN events from the literature, we have identified a
region of parameter space that defines that literature sample.

We applied this criteria to the SNLS archival data and discovered (or
recovered) three SLSN-I candidates within the redshift range of our
rate calculation, two of which have previously been identified. We
performed a Monte Carlo simulation of the SNLS to determine the rate
of SLSNe required in order for SNLS to detect these three events.  We
found the rate to be $\rho_{\mathrm{slsn}} =
91^{+76}_{-36}$\,SNe\,yr$^{-1}$\,Gpc$^{-3}$. This measurement is
consistent with what little was previously known about the rate of
SLSNe, and, when combined with other measurements, the redshift
evolution is consistent with that of the cosmic star-formation
history. We estimated the rate of ULGRBs based on the published
samples of these events and find their rate to be comparable with the
rate of SLSNe at a similar redshifts further demonstrating that these
events may be connected though a common progenitor.  We have also
studied the properties of the host galaxies associated with our SLSN
candidates, and find them to be consistent with the distribution of
the known sample of SLSNe.

\section*{Acknowledgments}

MS acknowledges support from EU/FP7-ERC grant no [615929].

This work is based on observations obtained with MegaPrime/MegaCam, a
joint project of CFHT and CEA/DAPNIA, at the Canada-France-Hawaii
Telescope (CFHT) which is operated by the National Research Council
(NRC) of Canada, the Institut National des Sciences de l'Univers of
the Centre National de la Recherche Scientifique (CNRS) of France, and
the University of Hawaii. This work is based in part on data products
produced at the Canadian Astronomy Data Centre as part of the
Canada-France-Hawaii Telescope Legacy Survey, a collaborative project
of NRC and CNRS.

Some of the data presented herein were obtained at the W.M. Keck
Observatory, which is operated as a scientific partnership among the
California Institute of Technology, the University of California and
the National Aeronautics and Space Administration. The Observatory was
made possible by the generous financial support of the W.M. Keck
Foundation.

\bibliographystyle{mnras}
\bibliography{Paper}

\bsp
\appendix

\section{SLSN parameter space}
\label{sec:slsn-parameter-space}

We choose an ellipsoid to define the parameter space of SLSNe in terms
of the three main fit parameters of the magnetar model,
$\tau_\mathrm{m}$, $B_{14}$ and $P_{\mathrm{ms}}$. A position, in
Cartesian coordinates, on a generic ellipsoid can be defined using
equation \ref{eq:Ellipsoid}.
\begin{equation}
\label{eq:Ellipsoid}
\left( \begin{matrix}
x \\
y \\
z
\end{matrix} \right)
=
\mathbf{A}
\left( \begin{matrix}
R_x\cos(u)\cos(v) \\
R_y\cos(u)\sin(v) \\
R_z\cos(v)
\end{matrix} \right)
+ \mathbf{C}
\end{equation}
where \textbf{A} is a rotation matrix, \textbf{R} is the radius of the
ellipsoid and \textbf{C} is its center. The following conditions must
be satisfied: $-\pi /2 \leq u \leq \pi /2$ and $-\pi \leq v \leq \pi$.
We have set up our parameter space with $\tau_\mathrm{m}$ along the
$x$-axis, $B_{14}$ along the $y$-axis and $P_{\mathrm{ms}}$
corresponding to the $z$-axis. Using the Khachiyan Algorithm
\citep{Aspvall19801,KHACHIYAN198053}, we have performed a fit to find
an ellipsoid that best describes the known population of SLSNe.  We
found the ellipsoid to have the following properties:
\begin{equation}
\label{eq:A}
\mathbf{A} =
\left( \begin{matrix}
-0.065 & -0.744 & -0.665 \\
 0.064 & -0.668 & 0.742 \\
-0.996 &  0.006 & 0.091
\end{matrix} \right)
\end{equation}

\begin{equation}
\label{eq:R}
\mathbf{R} =
\left( \begin{matrix}
R_x \\
R_y \\
R_z
\end{matrix} \right)
=
\left( \begin{matrix}
1.66 \\
2.62 \\
15.26
\end{matrix} \right)
\end{equation}

\begin{equation}
\label{eq:C}
\mathbf{C} =
\left( \begin{matrix}
29.65 \\
1.63 \\
2.61
\end{matrix} \right)
\end{equation}

\section{Light curves}
\label{sec:light-curves}
\begin{center}
\begin{table}
\label{table:07D3bs}
\caption{AB photometry of SNLS-07D3bs}
\begin{tabular}{|r|r|r|r|r|r|}
\hline
  \multicolumn{1}{c|}{MJD} &
  \multicolumn{1}{c|}{Phase} &
  \multicolumn{1}{c|}{g$_\mathrm{M}$} &
  \multicolumn{1}{c|}{r$_\mathrm{M}$} &
  \multicolumn{1}{c|}{i$_\mathrm{M}$} &
  \multicolumn{1}{c|}{z$_\mathrm{M}$} \\
\hline
54140.5 & -42.4 &  & 23.59  (0.07) & 23.33  (0.08) & 23.63  (0.19)\\
54144.5 & -38.4 & 23.42  (0.03) & 23.22  (0.04) & 23.04  (0.05) & \\
54151.5 & -31.4 & 22.70  (0.02) & 22.59  (0.03) & 22.57  (0.04) & 22.54  (0.11)\\
54153.5 & -29.4 & 22.58  (0.02) & 22.43  (0.02) & 22.41  (0.03) & \\
54172.6 & -10.3 &  &  & 21.62  (0.04) & \\
54176.4 & -6.5 & 22.02  (0.01) & 21.73  (0.01) & 21.62  (0.02) & 21.57  (0.04)\\
54179.5 & -3.4 & 22.07  (0.01) & 21.73  (0.01) & 21.59  (0.02) & 21.66  (0.03)\\
54183.5 & 0.6 & 22.17  (0.02) & 21.75  (0.01) & 21.60  (0.02) & \\
54186.4 & 3.5 & 22.22  (0.02) & 21.76  (0.02) & 21.56  (0.02) & 21.49  (0.03)\\
54197.5 & 14.6 &  & 21.94  (0.03) & 21.63  (0.02) & 21.59  (0.05)\\
54201.4 & 18.5 & 22.70  (0.02) & 21.99  (0.02) & 21.68  (0.03) & \\
54205.4 & 22.5 & 22.82  (0.02) & 22.05  (0.03) & 21.70  (0.02) & 21.62  (0.04)\\
54209.4 & 26.5 & 22.95  (0.02) & 22.18  (0.02) & 21.74  (0.02) & \\
54213.5 & 30.6 & 23.10  (0.06) & 22.28  (0.02) & 21.79  (0.02) & 21.69  (0.05)\\
54229.4 & 46.5 & 23.86  (0.07) & 22.71  (0.03) & 22.12  (0.02) & 21.85  (0.19)\\
54230.3 & 47.4 &  &  &  & 21.87  (0.07)\\
54233.3 & 50.4 &  &  & 22.11  (0.08) & \\
54234.4 & 51.5 & 24.05  (0.05) & 22.92  (0.03) & 22.26  (0.07) & \\
54251.4 & 68.5 &  & 23.63  (0.13) & 22.68  (0.05) & 22.16  (0.08)\\
54255.3 & 72.4 & 25.05  (0.34) & 23.65  (0.08) & 22.72  (0.09) & \\
54259.3 & 76.4 & 25.47  (0.14) & 23.92  (0.11) & 22.83  (0.03) & 22.42  (0.08)\\
54262.4 & 79.5 & 25.60  (0.20) & 23.90  (0.07) & 22.91  (0.10) & \\
54266.3 & 83.4 & 25.68  (0.35) & 24.20  (0.15) & 23.01  (0.04) & 22.40  (0.08)\\
\hline\end{tabular}
\end{table}
\end{center}

\newpage
\begin{figure}
\centering
\includegraphics[scale = 0.4]{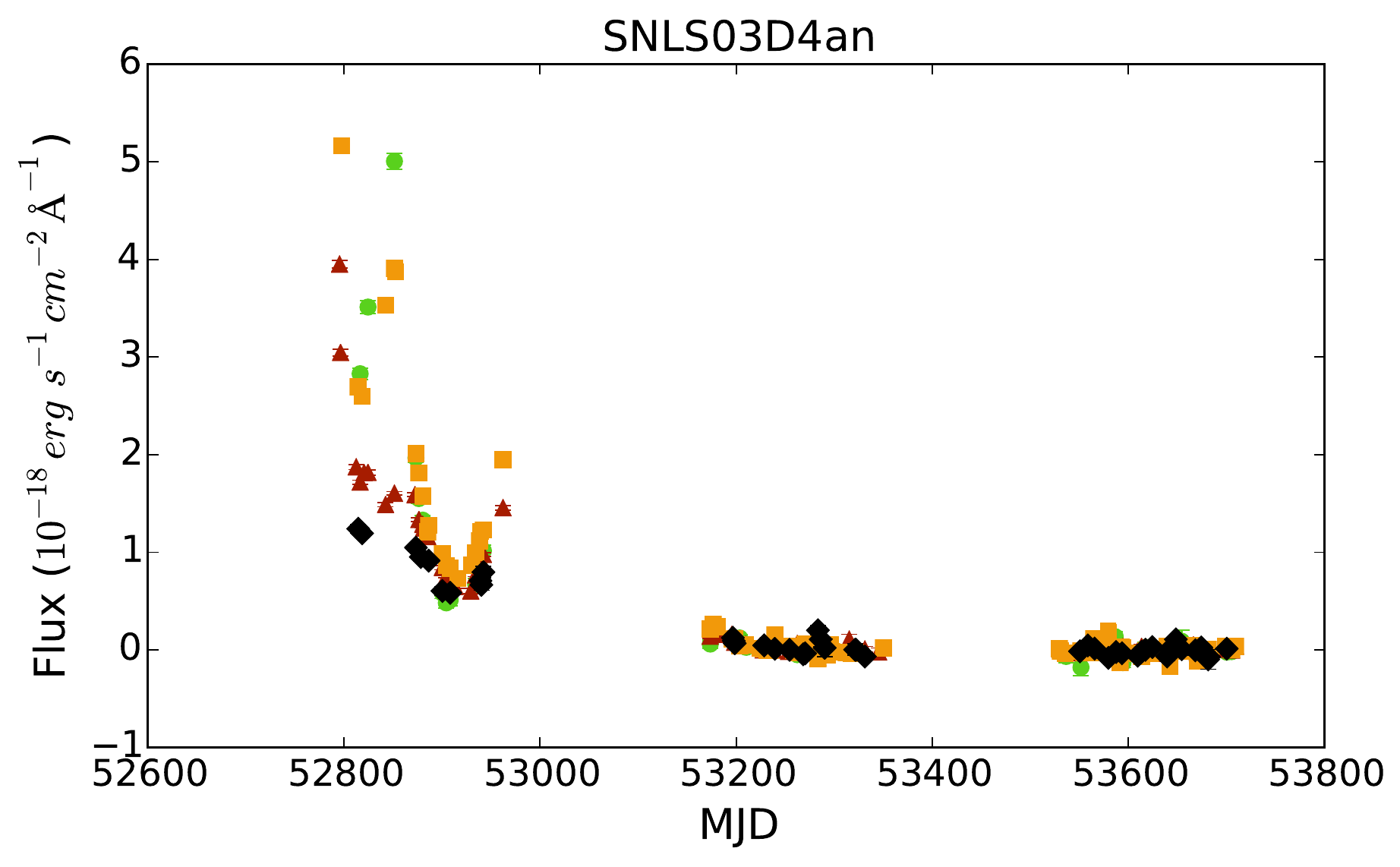}
\includegraphics[scale = 0.4]{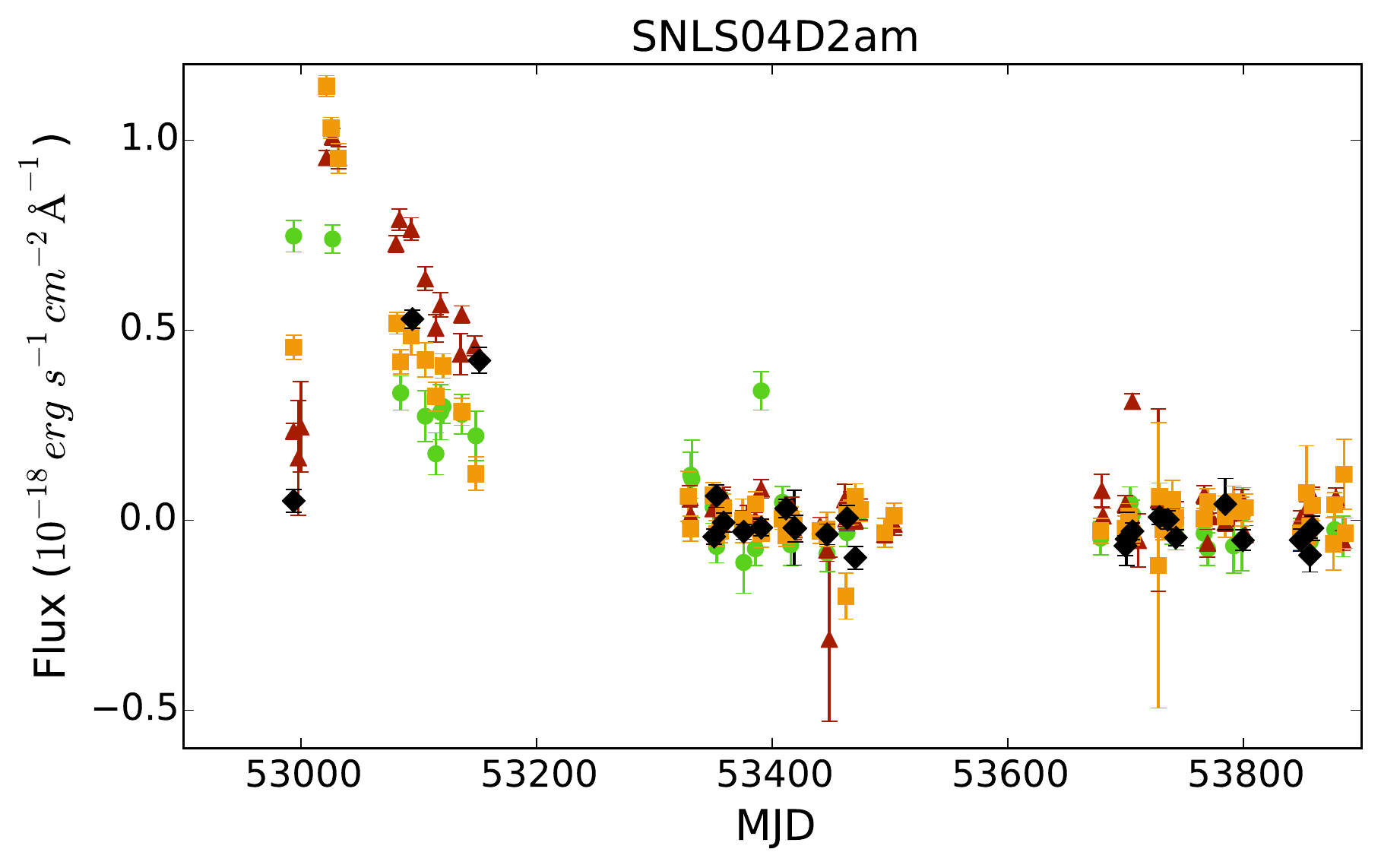}
\includegraphics[scale = 0.4]{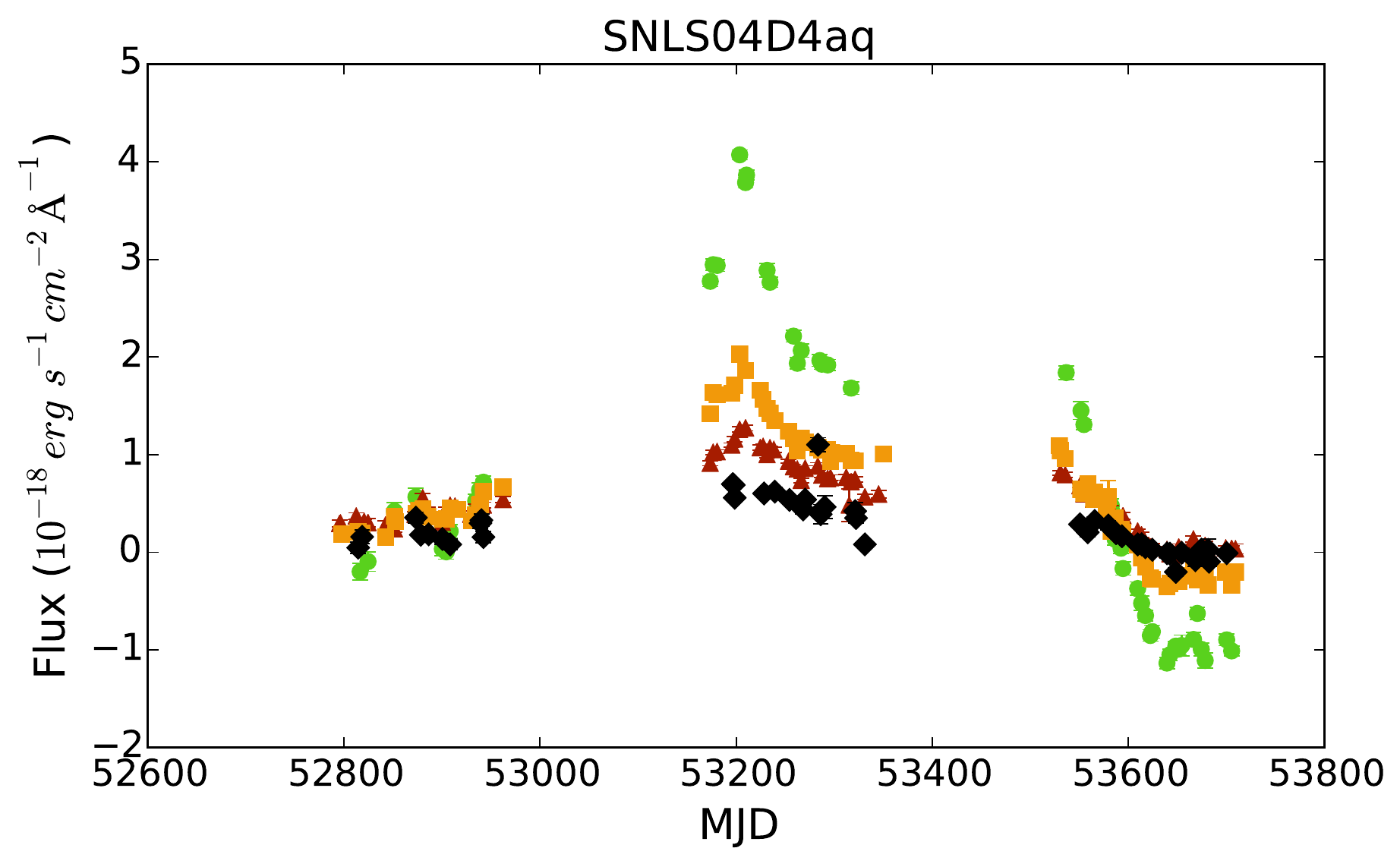}
\includegraphics[scale = 0.4]{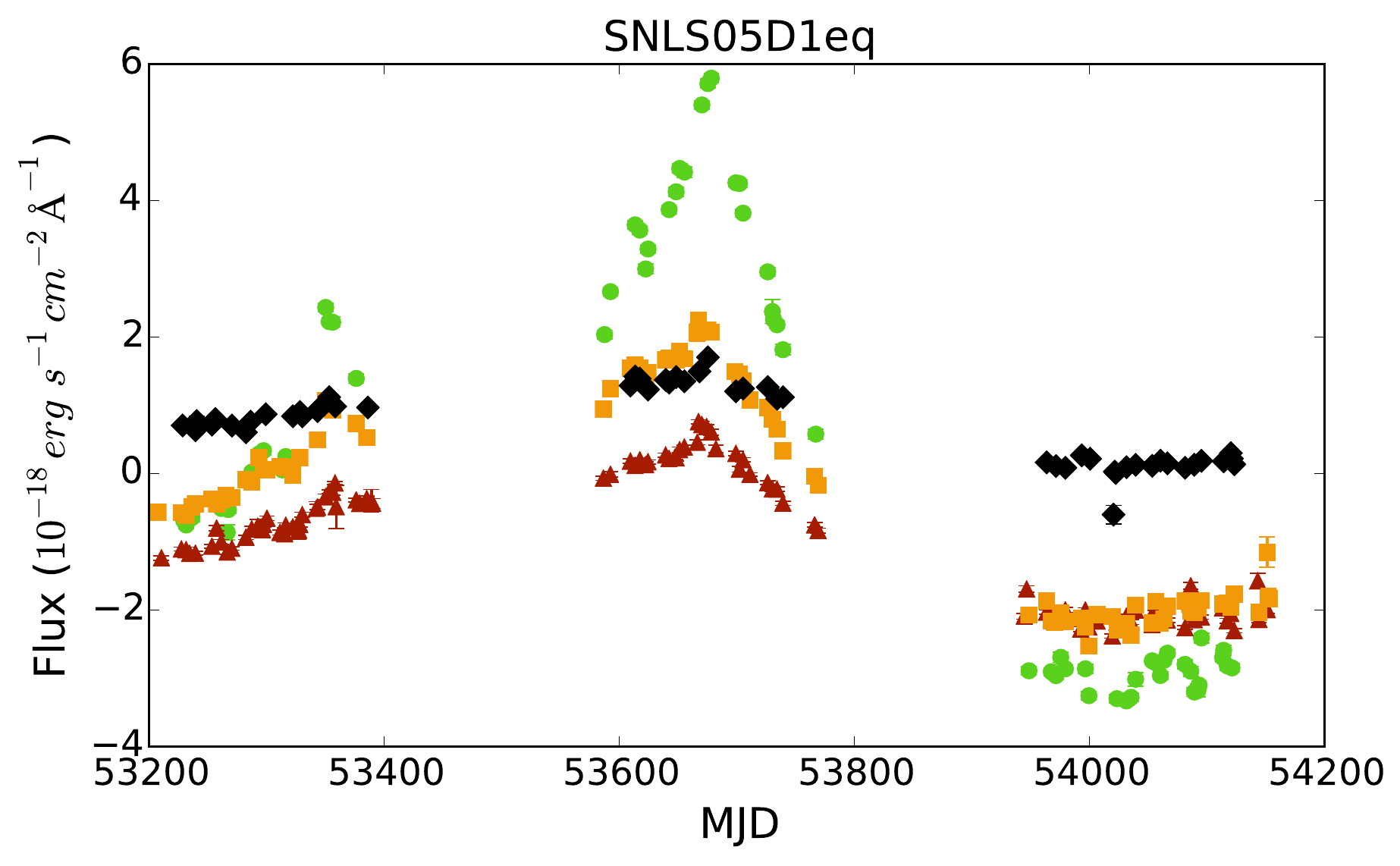}
\includegraphics[scale = 0.4]{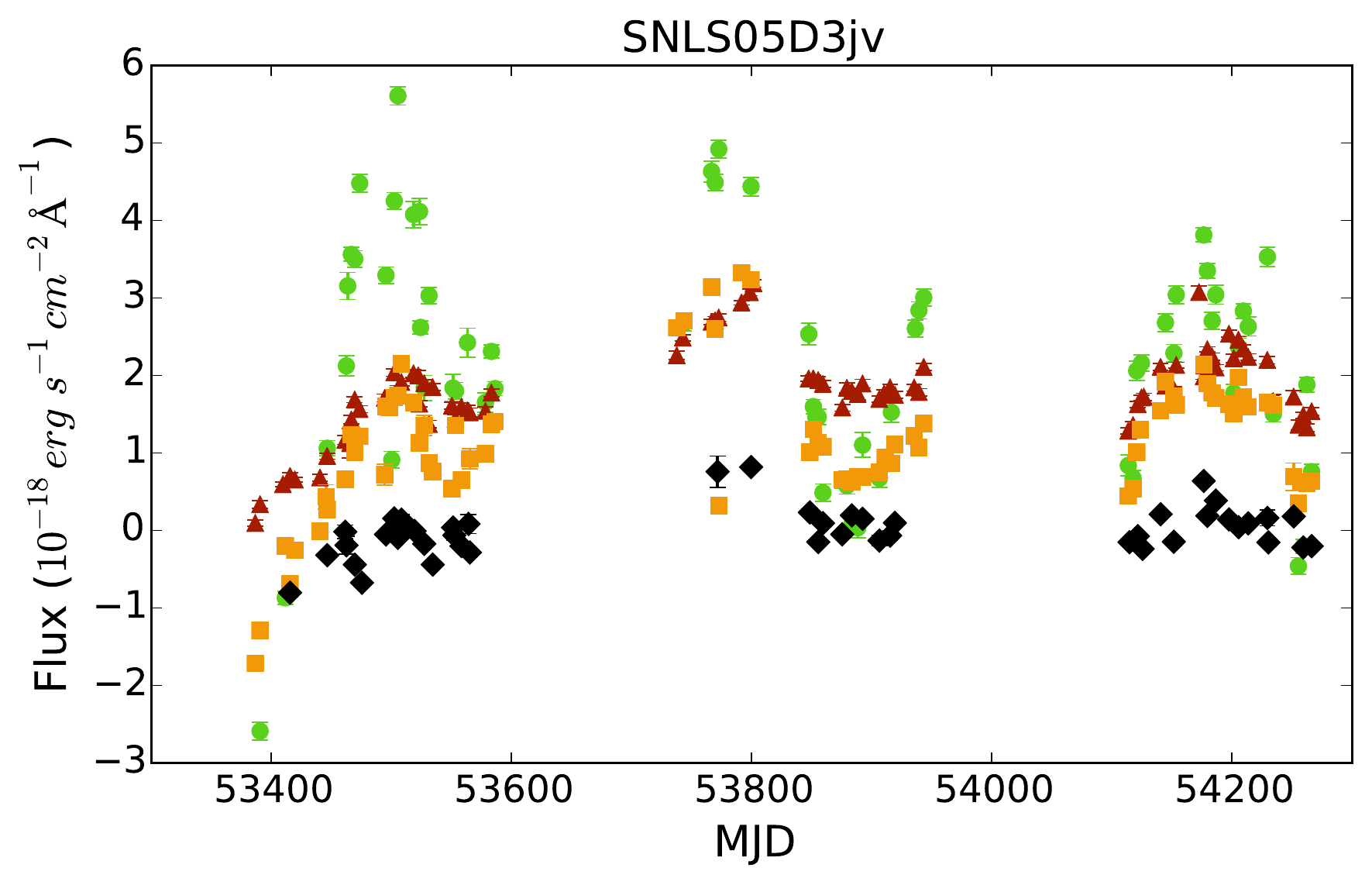}
\end{figure}

\begin{figure}
\centering
\includegraphics[scale = 0.4]{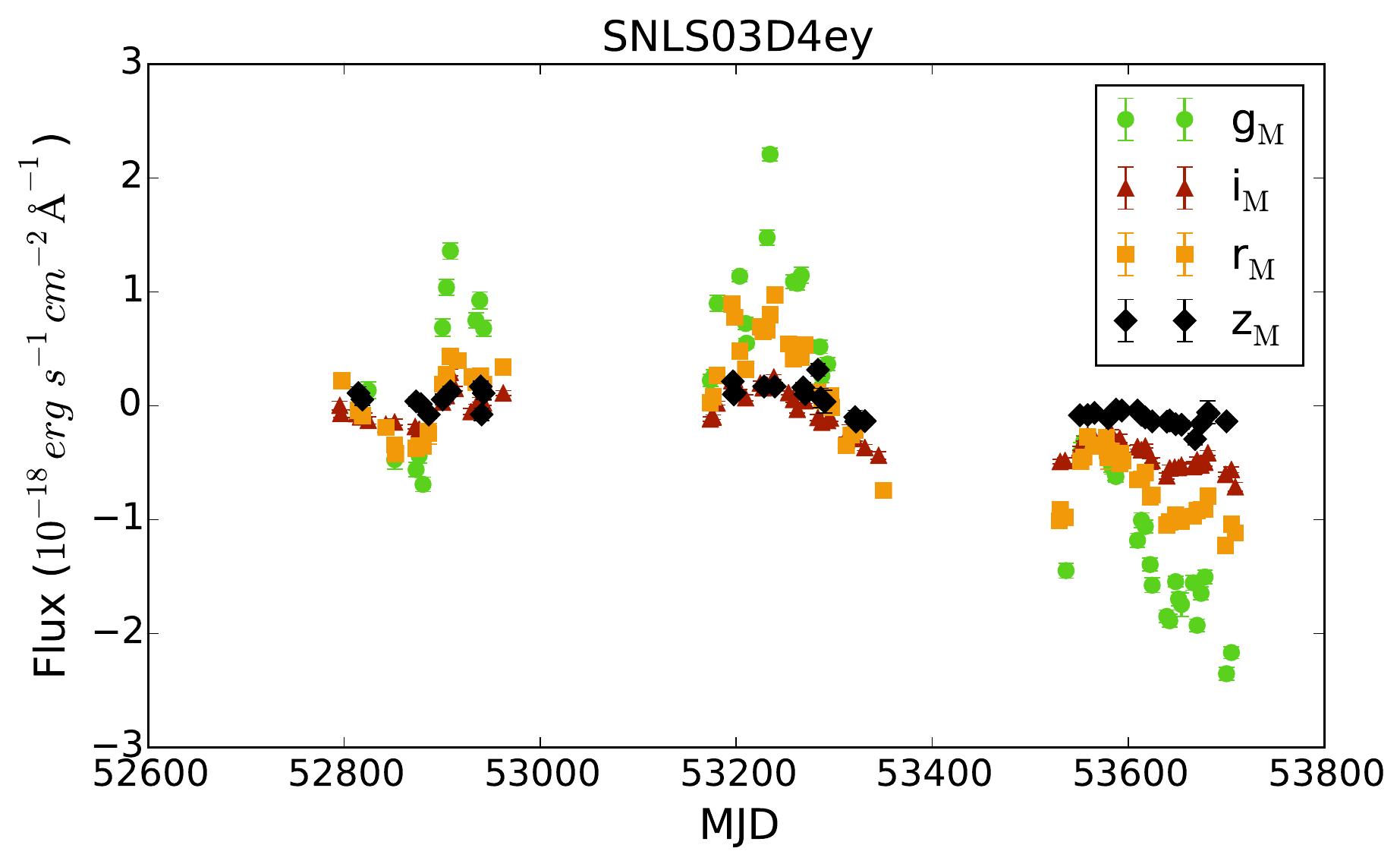}
\includegraphics[scale = 0.4]{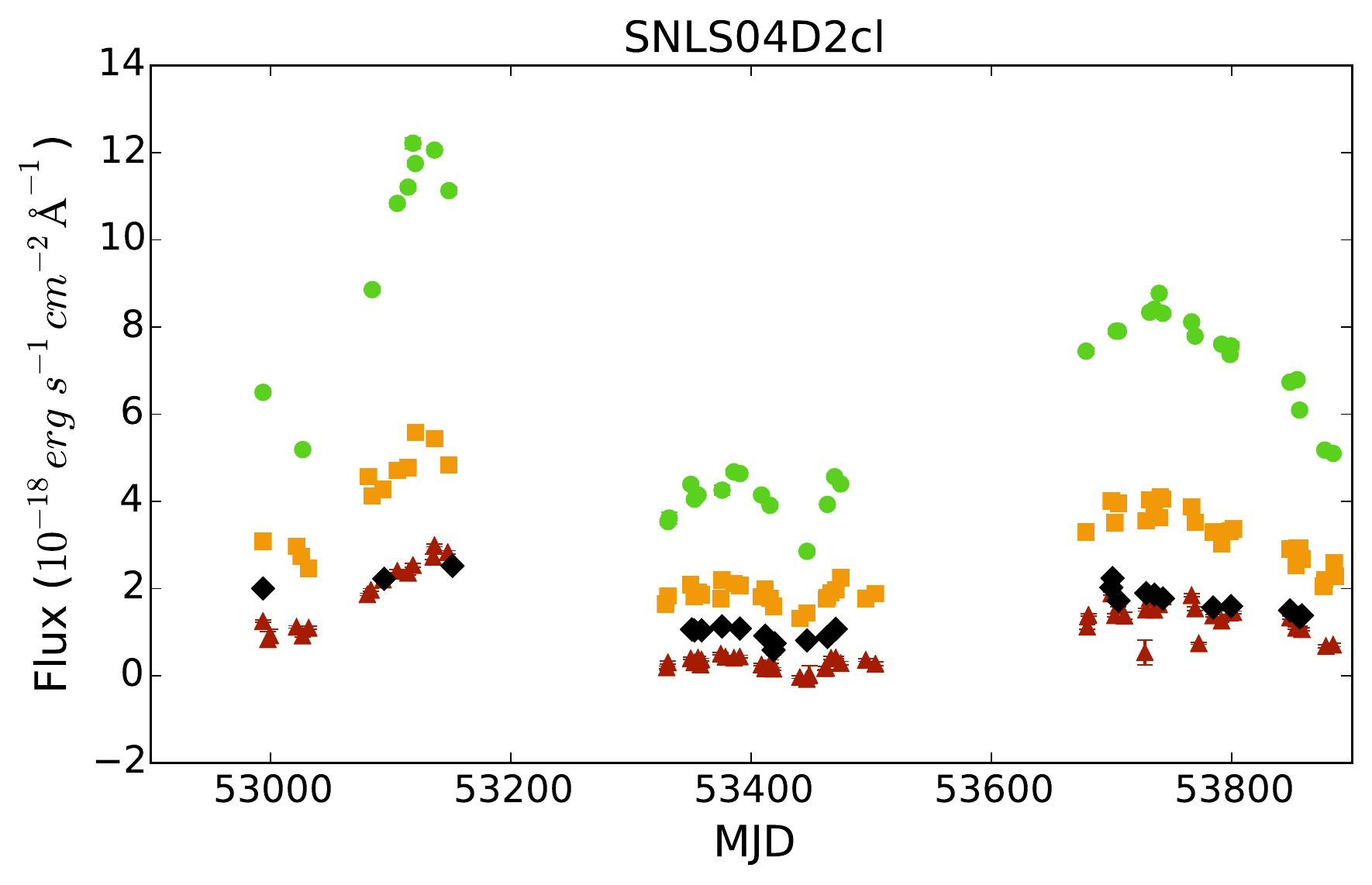}
\includegraphics[scale = 0.4]{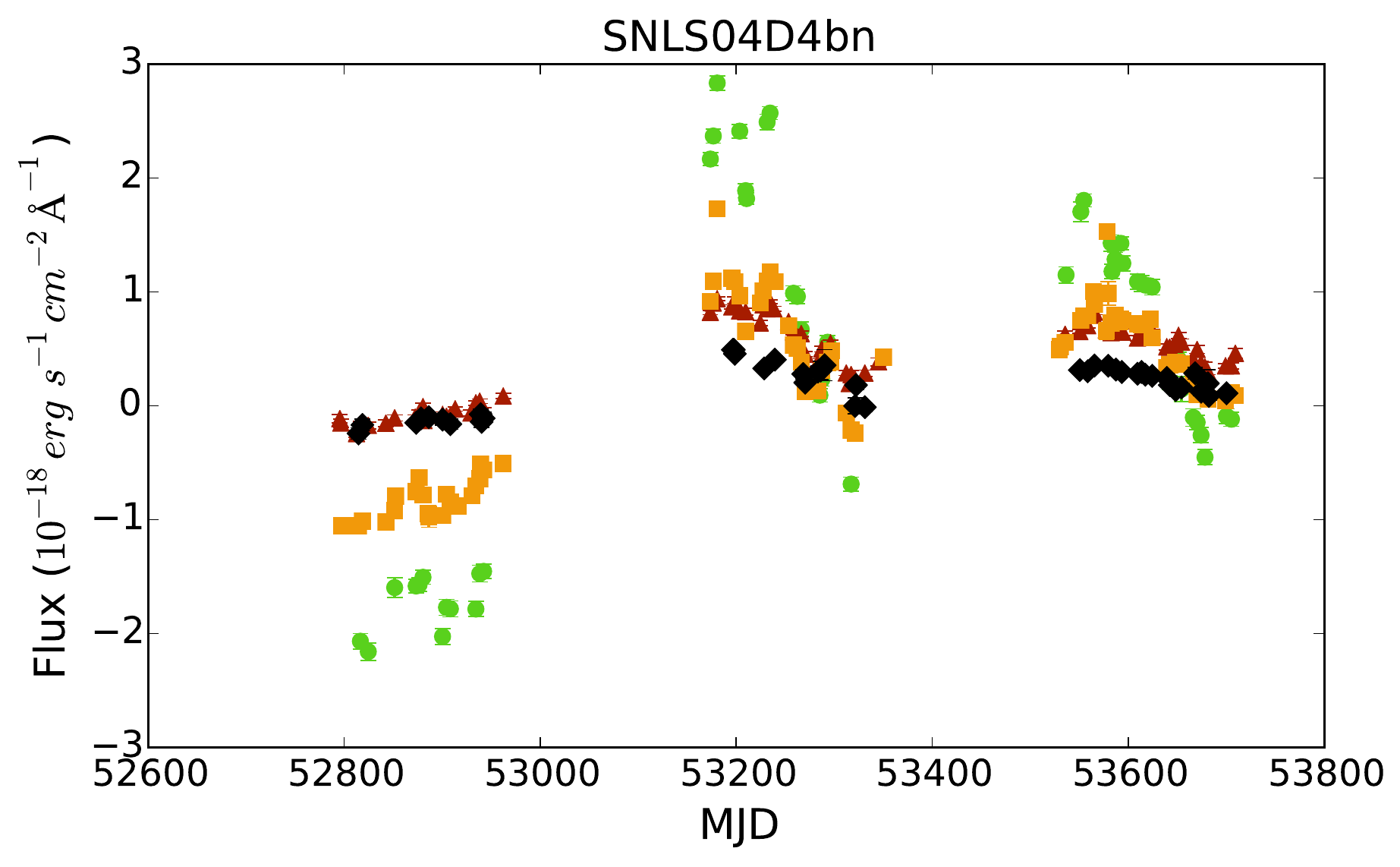}
\includegraphics[scale = 0.4]{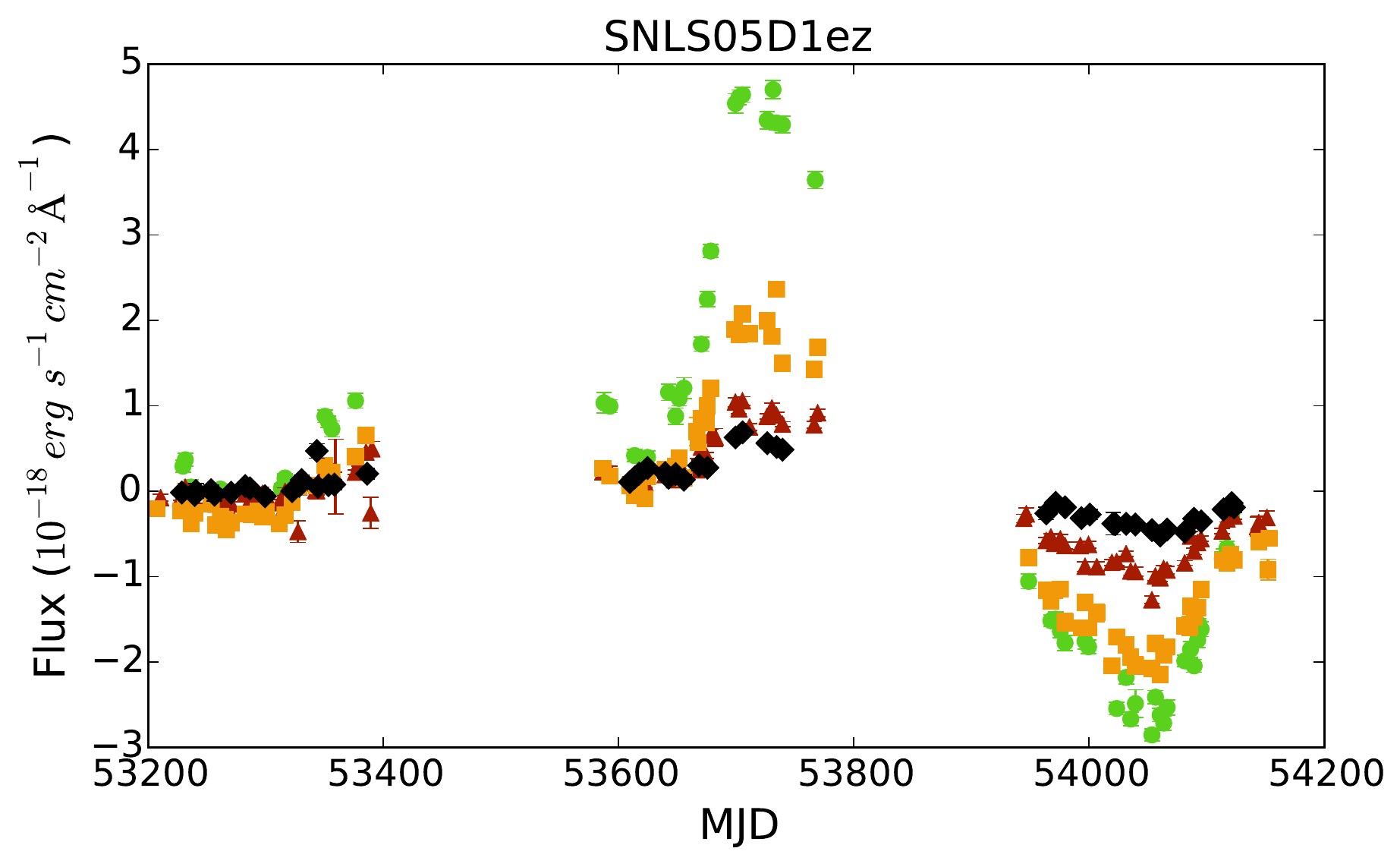}
\end{figure}

\begin{figure}
\centering
\caption{Three season light curves of SLSNe candidates found within the SLSN-I definition which do not pass our visual inspection. All events have 5$\sigma$ detections in multiple season and also show clear signs of multiple maxima and weak colour evolution.}
\end{figure}

\bsp
\label{lastpage}
\end{document}